\documentstyle[aps,psfig,preprint]{revtex}
\tighten

\newcommand{\ep}{\varepsilon}

\def\be{\begin{eqnarray}}
\def\ee{\end{eqnarray}}
\newcommand {\bP} {\bbox{P}}
\newcommand {\bp} {\bbox{p}}
\newcommand {\br} {\bbox{r}}
\newcommand {\bj} {\bbox{j}}
\newcommand {\bG} {\bbox{\ G}}
\newcommand {\bT} {\bbox{ T}}

\newcommand {\bV} {\bbox{ V}}
\newcommand {\bcV} {\bbox{\cal V}}
\newcommand {\bcT} {\bbox{\cal T}}
\newcommand {\bcG} {\bbox{\cal G}}
\newcommand {\bSigma} {\bbox{\Sigma}}
\newcommand {\bK} {\mbox{\boldmath$K$}}
\newcommand {\bk} {\mbox{\boldmath$k$}}
\newcommand {\bq} {\mbox{\boldmath$q$}}
\begin{document}
\title{A quantum kinetic equation for Fermi-systems  
                   including three-body correlations}
\author{Arm{e}n Sedrakian}
\address{Center for Radiophysics and 
Space Research 
and Department of Astronomy, \\Cornell University, Ithaca 14853, NY}
\author{Gerd R\"opke} 
\address{Department of Physics, Rostock University, 
18051 Rostock, Germany 
}
\maketitle
\begin{abstract}
A  single-time quantum transport equation, which includes effects beyond the  
quasiparticle approximation,  is derived  for Fermi-systems in the 
framework of non-equilibrium real-time Green's functions theory. 
Ternary correlations are incorporated in  the kinetic description
via a cluster expansion for  the self-energies  
(e.g., the transport vertex and the width) 
truncated at the level of three-body  scattering 
amplitudes. A finite temperature/density formulation of the 
three-body problem is given. Corresponding three-body equations 
reduce to the well-known Faddeev equations in the vacuum limit.
In  equilibrium the equation of state contains 
virial corrections proportional to the third quantum virial coefficient. 
\end{abstract}

\section{Introduction}
\noindent
A reliable description of the dynamics of Fermi-systems is commonly 
achieved in the quasiclassical (QC) limit, which is based on the concept 
of a classical particle moving in the mean field and along a 
classical trajectory between successive instantaneous collisions.
The QC picture assumes  a hierarchy of  time-scales; the 
time-scales on which  the renormalization of the 
excitation spectra and collisions occur are expected to be smaller 
than the mean time-scale between  encounters of dressed quasiparticles.
A similar assumption applies to the length-scales.  
As well known, such a separation is possible in the limiting cases 
of a classical dilute gas and a highly degenerate  
Fermi-liquid (for a review see,  e.g., ~\cite{BP,LP}). 
The dynamical information in quasiclassical approximation (QCA)
is contained in the single particle distribution function; the 
higher order distributions (determined form the BGKKY hierarchy) 
appear as time-dependent functionals of single particle distribution
(for a review see, e.g., ~\cite{ZMR}). 
A second element of kinetic description of 
Fermi-systems is the quasiparticle approximation (QPA), which assumes 
a sharp  functional dependence of the energies of the quasiparticles
on their momenta. The QPA is sufficient to constrain the time-scale
(but not the length-scale) hierarchy to the QC limit;
it does not, however, appear to be a necessary condition.

The need of extension of kinetic description of Fermi-systems beyond 
QPA arises in a number of physical situations.
An example is the  excited nuclear matter produced in heavy-ion 
collision experiments at energies $\sim 100$~Mev/nucleon
(for a review see, e.g., ~\cite{BERTSCH}).
The expanding nuclear system traverses a broad region in the
density-temperature plane and, in particular, the temperature could be of
order the Fermi temperature.  In such  situations 
the spectral function of the single particle excitations acquires a 
finite width. If the half-width of the spectral function
is small compared with the amplitude, the corrections to the 
QP picture are small, and the hierarchy of the time-scale still
holds (in other words, a distinction can be made between the renormalization 
effect leading to the broading of the QP peak and the elementary collision 
events between the single particle excitations with renormalized spectra). 
The finite width of the spectral function introduces in the system   
correlations {\it beyond} the renormalization of the single particle spectrum,
which lead to the so-called virial corrections,
and in particular  to formation of 
bound states (in nuclear collisions - deuterons~\cite{BALDO,DB,RS}, tritons, 
alpha particles, etc.) and off-mass-shell propagation (such as 
Landau-Pomeranchuk-Migdal effect in the particle radiation 
processes ~\cite{VOSKRESENSKY}).
Situation, similar to the one described above, 
is encountered also in the astrophysical context;
e.g. in the finite temperature isospin asymmetical 
nuclear matter in the 
supernova collapse and proto-neutron stars.

The picture above suggests seeking a kinetic equation for 
Fermi-systems that accommodates the separation of the time and length-scales 
(and hence the QC approximation)  while allowing for effects beyond the 
QPA. [Formally this means that the time-scales
of variation of the self-energy functions (providing the renormalization 
and transport vertex) are small compared to the evolution time-scales
for the propagators related to the quasiparticle distribution function].

Initial studies of the virial corrections to the Boltzmann equation were 
carried out using the density matrix approach ~\cite{COHEN}; progress
in this direction is represented by Snyder's kinetic equation~\cite{SNYDER}, 
where  Bogoliubov's ansatz of the weakening of the initial 
correlations is used to close the BGKKY hierarchy at the two-particle 
level. The partition of the virial corrections between the collision 
integral and the drift term is not unambiguous~\cite{LALOE}, 
however the kinetic equation reproduces the correct form of the lowest order  
(second) virial coefficient ~\cite{SNYDER}.
The formation of the bound states and three-body processes within 
the density matrix approach has been considered e.g. 
by McLennan~\cite{MCLENNAN}, Klimontovich et al ~\cite{KLIMONTOVICH}
and R\"opke and Schulz~\cite{RS}, who derive coupled 
kinetic equations for a gas viewed as a reacting mixture of atoms 
and diatomic molecules or nucleons and their bound states
(the authors do not consider the third virial coefficient, however).

The transport theory, which is based on the real-time Green's functions 
formalism and  is largely due to  Kadanoff and Baym  ~\cite{KB} and  
Keldysh~\cite{KELDYSH}, provides an alternative to the density matrix approach,
and allows systematic diagrammatic approximations to the propagators and 
self-energies. 
The reduction scheme for the double-time integro-differential 
Kadanoff-Baym equations for the Green's 
functions to a single-time  Boltzmann-type kinetic uses the QCA, which, as discussed above, 
separates the time and length-scales associated with 
individual collisions from those characterizing the inter-collision 
dynamics, and the QPA, which assumes 
a delta-shaped spectral function. 
The form and further approximations for the scattering-in  
and -out rates, represented by the Feynman diagrams of elementary scattering
processes, vary considerably depending on the specifics 
of the dynamics and composition of the  
underlying system; cf., e.g., ~\cite{SERENE,DAN,ARONOV,BMALF}.

The quantum kinetics beyond QPA has been addressed in a number of recent 
works~\cite{BMALF,KMGR,SLM,SL}. 
The memory effects in the  non-Markovian kinetic 
equations, as shown in ~\cite{KMGR}, provide a sufficiently 
general description of correlations in quantum systems; 
the second order virial corrections emerge
when the time-retarded spectral function is
expanded to the next-to-leading order in retardation. 
This form of second-order quantum 
virial correction agrees  with the equilibrium
result derived for semiconductors~\cite{ZIM}, Coulomb
plasmas ~\cite{KKL} and finite temperature nuclear matter ~\cite{SRS}.
The second order virial corrections in the  local-time description 
and for the electron-impurity system have been given in ~\cite{SLM} 
in terms of the next-to-leading  order expansion of the spectral function with 
respect to the width ~\cite{SL}.

The solution of the 
three-body problem for an isolated system is well know, and is represented 
by the Faddeev~\cite{FADDEEV} or AGS ~\cite{AGS} equations.
The Kadanoff-Baym formalism has been applied to derive kinetic equations 
including triple collisions by  Bezzerides and DuBois~\cite{DUBOIS};
their study, however, has been restricted mainly to the Born 
approximation, and consequently the collision integrals 
contain singularities due to the 
non-Fredholm type kernels of the scattering amplitudes. 
A different motivation for studying the three-body equations
comes from the problem of the imaginary nuclear potentials, where the  
2-particle-hole channel dominates 
ground state correlations in nuclei ~\cite{SCHUCK}. A self-consistent random 
phase approximation to the respective three-body problem at zero-temperature 
has been given in ref. ~\cite{SCHUCK}, while the the finite temperature 
discussion  of the latter problem is contained
in refs. ~\cite{DAN,DS} as an example of the 
application of the cutting rules for the 
multi-contour ordered Green's functions due to Danielewicz ~\cite{DAN}.
In the near-equilibrium situation the nucleon-deuteron system has been studied
at finite temperatures in the  AGS formulation in refs.  ~\cite{BRS,BR}.

Despite the progress mentioned above, the subject of the kinetics with 
three-body correlations needs further attention
since (i)~one would like to arrive at a kinetic 
equation which in the equilibrium limit leads to third order quantum virial corrections 
to the equation of state, 
(ii)~the three-body collision processes should be free of 
singularities, therefore the appropriate Faddeev or AGS amplitudes should 
be used in the collision integrals (see also ~\cite{BR}); 
(iii)~the scattering amplitudes in the medium should 
reduce to the three-body Faddeev equations which are exact 
in the vacuum limit.

The purpose of this paper is to address these items 
and to  provide a link between the kinetic description beyond 
the QPA and three-body correlations. 
We adopt the familiar methodology of next-to-leading order 
expansion of the spectral function with respect to the spectral 
width, however instead of 
the decomposition of Green's functions into 
pole and off-pole terms~\cite{KMGR,SLM,SL}, we use a 
decomposition in powers of the  width. 
In the introduction to  Section 2 we discuss the 
double-time kinetic equation and the 
gradient approximation to this equation (readers familiar with the 
formalism may skip this part of the section; the notation is as in 
Kadanoff and Baym~\cite{KB}). Further subsections introduce the approximations
to the spectral function, the method of decomposition of the diagrams in the 
QP and correlated parts, the Kadanoff-Baym ansatz and, finally, the
single time kinetic equations. Section 3 derives the three-body equations
at finite temperatures and densities. The transport vertex and the collision 
integrals are discussed in Section 4.
The equilibrium limit of the kinetic regime 
and the third quantum virial coefficient are derived in Section 5. 
Section 6 contains  a brief summary of our results.

\section{Dyson equations}
Consider a non-relativistic 
Fermi-system interacting via two-body forces. The 
Hamiltonian of the system reads
\be\label{1} 
H&=&\frac{1}{2m}\sum_{\sigma}\int d^4 x \vec \nabla\psi_{\sigma}^{\dagger}(x)
\vec \nabla\psi_{\sigma}(x) \nonumber\\
&+&\frac{1}{2}\sum_{\sigma\sigma'}
\int d^4x d^4x' \psi_{\sigma}^{\dagger}(x)
\psi_{\sigma'}^{\dagger}(x') 
V_{\sigma,\sigma'}(x,x')
\psi_{\sigma'}(x') \psi_{\sigma}(x),
\ee
where $\psi_{\sigma}(x)$ are the Heisenberg field operators, $x=(\vec r, t)$ 
is the space-time four vector, $\sigma$ stands for the internal 
degrees of freedom (spin, isospin, etc.). 
We shall use  the matrix form 
of arrangement of  time-contour ordered real time Green's functions, which is
particularly suited when initial correlations are absent.
In this case  the upper branch, $c_-$,  
runs from $-\infty$ to $\infty$ and the 
lower branch, $c_+$, from $\infty$ 
to $-\infty$. The initial correlation are commonly incorporated 
via the imaginary-time piece of the contour which accounts for 
the Kubo-Martin-Schwinger boundary condition ~\cite{DAN}. 
We shall neglect this piece in the further consideration.
The single-particle  propagator matrix is defined  as
\be\label{2}
{\bG}_{12} &=&\left(\begin{array}{cc}  G^{c}_{12}& 
 G^{<}_{12}\\ 
 G^{>}_{12}&  {G}^{a}_{12}\end{array}\right) \nonumber \\
&=&\left(\begin{array}{cc} -i\langle T\,\psi_{\sigma_1}(x_1)\, 
\psi^{\dagger}_{\sigma_2}(x_2)\rangle
& i\langle \,  \psi_{\sigma_2}^{\dagger}(x_2)\, 
\psi_{\sigma_1}(x_1)\rangle\\ -i\langle \, \psi_{\sigma_1}
  (x_1)\,\psi_{\sigma_2}^{\dagger}(x_2)\rangle & -i\langle 
\tilde T\, \psi_{\sigma_1}(x_1)\, 
\psi_{\sigma_2}^{\dagger}(x_2)\rangle\end{array}\right), 
\ee
where the indexes $1,2... $ collectively denote the variables 
$(x_1, \sigma_1), (x_2, \sigma_2)...$,  the averaging is over 
a nonequilibrium state of  the system; $T$ and $\tilde T$ 
are chronological and anti-chronological time ordering operators, 
respectively.

The perturbative expansion for the  Green's functions ${\bG}$
can be arranged, as well known, 
in a manner similar to the ground state Feynman 
diagrammatic expansion utilizing Wick's theorem. 
The single-particle propagator, therefore,  obeys the Dyson equation
\be\label{DYSON}
{\bG}_{12}={\bG}_{12}^{(0)}+ {\bG}_{14}^{(0)}\otimes {\bSigma}_{43}\otimes{\bG}_{32} 
 ={\bG}_{12}^{(0)}+ {\bG}_{14}\otimes {\bSigma}_{43}\otimes{\bG}_{32}^{(0)} ,
\ee
where the superscript $(0)$ refers to the free propagator, $\otimes$ stands
for  matrix multiplication along with the folding over the internal (repeated) 
indexes and the self-energy matrix ${\bSigma}$ has 
a structure identical  to eq. (\ref{2}). 
The integro-differential form of the Dyson equation is obtained 
by acting the evolution 
operator ${\bG}_{01}^{-1}= i\partial/\partial t_1+\nabla^2_1/2m$ 
on eq. (\ref{DYSON}) (the energy scales are measured relative to 
the fermion chemical potential) and employing  
the free-particle Dyson equation, 
${\bG}_{01}^{-1}\otimes{\bG}_{12}^{(0)}=\bbox{\sigma}_z\,\,
\delta^{(4)}(x_1-x_2)$; here and below $\bbox{\sigma}_i$ ($ i 
=x, y, z$) are the  Pauli matrices and the subscript 0 refers to the
evolution operator of free particle. 
Subtracting the resulting equation from its conjugate one finds
\be\label{MATRIXKIN}
\left({\bG}_{02}^{-1\, *}-{\bG}_{01}^{-1}\right)\otimes
{\bG}_{12} = 
{\bG}_{13}\otimes{\bSigma}_{32}\otimes\bbox{\sigma}_z
-\bbox{\sigma}_z\otimes {\bSigma}_{13}\otimes{\bG}_{32}.
\ee 
The matrix structure of the Green's functions and the self-energies is
not the optimal one yet.  The retarded and advanced functions 
are preferable ~\cite{KELDYSH,DAN,DUBOIS}, 
since they obey an integal equation in the lowest order of 
QCA. In terms of real parts of these functions (${\rm Re}\, G^R = 
{\rm Re}\, G^A \equiv {\rm Re}\,  G$ and  ${\rm Re}\, \Sigma^R = 
{\rm Re}\, \Sigma^A \equiv {\rm Re}\,  \Sigma$)
the kinetic equation, e.g.,  for the $G^<$ component is
\begin{eqnarray}
\label{KINEQ2}
&&\left[  G_{03}^{-1} -{\rm Re} \Sigma_{13}, G_{32}^{<}\right]
- \left[{\rm Re}\,   G_{13},\Sigma_{32}^{<}\right] =
\frac{1}{2}\,\left\{  G_{13}^{>},  \Sigma_{32}^{<}\right\} 
-\frac{1}{2}\,\left\{  \Sigma_{13}^{>}, G_{32}^{<}\right\},
\end{eqnarray}
where $\left[~,~\right]$ and  $\left\{~,~\right\}$ denote the commutator
and anti-commutator, respectively,  and a summation (integration) over 
repeated indexes is assumed. If the 
dynamics of the system
permits a quasiclassical treatment, i.e., if
the characteristic inter-collision  length-scales are much greater than the 
inverse momenta and the relaxation times are much larger than 
the inverse frequencies, QCA can be applied by separating the slowly varying 
center-of-mass four-coordinates from the rapidly varying relative coordinates. 
Performing a Fourier transform with respect to the 
relative coordinates and keeping the first order gradients in the slow 
variable the QC Kadanoff-Baym  kinetic equation reads
\be
\label{KB} 
i\left\{{\rm Re}\, G^{-1}, \,   G^{<}\right\}_{\rm\small P.B.} 
+ i \left\{  \Sigma^{<}\,
,{\rm Re}\,   G \right\}_{\rm  P.B.}=  \Sigma^{<}G^{>} - \Sigma^{>}  G^{<} ,
\ee
where all functions depend on the four energy-momentum vector $p\equiv 
(\omega ,\,{\bp})$, and the center of mass space-time vector  $x$.
Here the four-component Poisson bracket is defined as
\be 
\{ f,\, g \}_{\rm\small  P.B.}
\equiv \frac{\partial f}{\partial \omega}\frac{ \partial g}{ \partial t} -
\frac{\partial f}{\partial t} \frac{\partial g}{\partial \omega} 
-\frac{\partial f}{\partial \bp}\frac{ \partial g}{ \partial \br} 
+ \frac{\partial f}{\partial \br} \frac{\partial g}{\partial \bp}. 
 \ee
Equation (\ref{KB}) is the starting point of the reduction of double time
kinetic equation to the single time  form;
one recognizes the left-hand-side as a generalized 
drift term, while the right-hand-side as the gain and loss 
terms of the collision integral.

\subsection{Spectral decomposition}

The Dyson equation for the retarded/advanced Green's functions
to the leading order in QCA  has an integral form, with the 
formal solution
\be\label{GR} G^{R/A}(p, x)
= \left[\omega -\epsilon_p-\Sigma^{R/A}(p, x) \pm i\eta \right]^{-1},
\ee
where the free single particle spectrum  is $\epsilon_p = p^2/2m$. 
The respective 
single particle spectral function, defined as $a(p,x) 
= i\left[G^R(p,x) - G^A(p,x)\right]$,  has the general form 
\be\label{SF} 
a(p,x) = \frac{\gamma(p,x)}{\left[\omega -\varepsilon_p(x)\right]^2
+\left[\gamma(p,x)/2\right]^2},
\ee 
where $\gamma(p,x) =i\left[ \Sigma^R(p,x) - \Sigma^A(p,x)\right]$ is the width  
of the spectral function and 
$\varepsilon_p(x) = \epsilon_p + {\rm Re}\,  \Sigma (p,x)\vert_{\omega = 
\ep_p}$ 
is the dispersion (the pole value of the retarded Green's function (\ref{GR})).
Following refs. ~\cite{ZIM,KKL,SRS}, we expand the spectral function 
up to  the next-to-leading order term in the power series expansion with
respect to the width $\gamma$ 
\be
\label{exp} 
a(p,x) &\simeq& 2\pi z(\bp, x)\delta(\omega -\ep_p(x)) -\gamma(p, x) 
\, \frac{{\cal P'}}{\omega -\ep_p(x)} ,  
\ee
with  the short-hand notation ${\cal P'}/(\omega -\ep_p(x))\equiv 
\partial/\partial  \omega~[ {\cal P}/(\omega -\ep_p(x))]$. Note that 
the self-energy appearing in the  
denominator of the second term of eq.~(\ref{exp}) via the dispersion relation 
is restricted, to first order in $\gamma$, to the mass-shell.
The wave-function renormalization, $z(\bp, x)$, in the same approximation is 
\be 
\label{Z} z(\bp, x) 
=1+\int\!\frac{d\omega'}{2\pi} 
\gamma(\omega',\bp,x)\, 
\frac{\cal P'}{\omega '-\omega}\Big|_{\omega=\ep_p}, 
\ee 
where we used the integro-differential form of the Kramers-Kronig relation:
\be 
\frac{\partial}{\partial \omega}{\rm Re}\, \Sigma(\omega,\bp, x)&=&
\int\!\frac{d\omega'}{\pi}  {\rm  Im}\, \Sigma(\omega',\bp, x)
\frac{{\cal P'}}{\omega-\omega'}=-\int\!\frac{d\omega'}{2\pi}  
\gamma(\omega',\bp, x)\frac{{\cal P'}}{\omega-\omega'}.\nonumber
\ee 
One may observe, combining expansion (\ref{exp}) and 
(\ref{Z}), that the spectral sum rule, 
\be \label{sum_rule}
\int\!\frac{d\omega}{2\pi}~a(p,x) = 1,
\ee
is fulfilled to any order in the expansion with respect to the width.

%%%%%%%%%%%%%%%%%%%%%%%%%%%%%%%%%%%%%%%%%%%%%%%%%%%%%%%%%%%%%%%%%%%%
\subsection{Correlation bracket}
The reduction of the  double-time kinetic description to a tractable
single time description requires projecting 
out different energy states, yet keeping the maximal information 
on the spectral properties of the system. The QPA projects out just a 
single energy value (defined by the on-mass-shell condition)
leaving out the off-mass-shell dynamics of the system. On the other hand,
 keeping the full spectral
function in the kinetic scheme is prohibitive for numerical applications,
and is not necessary if the off-mass-shell effects appear as corrections. 
The process of projecting out the energy states can be systemized in terms 
of simple decomposition rules (see also ~\cite{ZIM,KKL,SRS}). 
Consider an auxiliary step of decomposition of $\omega$-integrated 
arbitrary off-mass-shell function(s) with respect to the 
spectral width. A substitution of the spectral function in the integral
allows us to establish simple rules; e.g. for a function  of a single energy argument
one finds
\be 
\int a(\omega) F(\omega)d\omega 
&=& F(\varepsilon_p) + \left\{ F(\varepsilon_p) \right\}_{\rm corr}, 
\nonumber \\
\left\{ F(\varepsilon_p) \right\}_{\rm corr}&=&
\int\!\frac{d\omega}{2\pi}
\gamma(\omega,\bp,x)\,
\frac{\cal P'}{\varepsilon_{p} -\omega}
 [F(\omega) -  F(\varepsilon_p)],
\ee
where ``corr'' stands for correlation.
A product of two functions of a single energy argument decomposes as
\be
\int a(\omega) F_1(\omega)\, F_2(\omega)d\omega
&=& F_1(\varepsilon_p)\, F_2(\varepsilon_p) + 
F_1(\varepsilon_p)\, \left\{ F_2(\varepsilon_p) \right\}_{\rm corr}
\nonumber \\
&+& F_2(\varepsilon_p)\left\{ F_1(\varepsilon_p) \right\}_{\rm corr} 
+\left\{ F_1(\varepsilon_p)\, F_2(\varepsilon_p)  \right\}_{\rm corr},
\ee
\be 
\left\{ F_1(\varepsilon_p)\, F_2(\varepsilon_p)  \right\}_{\rm corr}
=\int\!\frac{d\omega}{2\pi}
\gamma(\omega,\bp,x)\,
\frac{\cal P'}{\varepsilon_{p} -\omega}
 [F_1(\omega) -  F_1(\varepsilon_p)]
 [F_2(\omega) -  F_2(\varepsilon_p)].
\ee
For a single function of two energy arguments one finds
\be 
\int a(\omega_1) a(\omega_2) F(\omega_1, \omega_2)\, d\omega_1\, d\omega_2
 = F(\varepsilon_{p_1}, \varepsilon_{p_2})
+\frac{1}{2}\left(\left\{ F(\varepsilon_{p_1},\ep_{p_2}) \right\}_{\rm corr}
+\left\{ F(\varepsilon_{p_2}, \ep_{p_1})\right\}_{\rm corr}\right),
\ee
\be 
\left\{ F(\ep_{p_1}, \ep_{p_2}) \right\}_{\rm corr}
=\int\!\frac{d\omega_1}{2\pi}\gamma(\omega_1,\bp,x)\,
\frac{\cal P'}{\varepsilon_{p_1} -\omega_1} 
[F(\omega_1, \varepsilon_{p_2})-F(\varepsilon_{p_1},\varepsilon_{p_2})],
\ee
and so on. 
Any diagram now can be decomposed in terms of these rules in
the leading and next-to-leading order terms in the spectral 
width. The coherent terms of the type $\{F_1F_2\}_{\rm corr}$ assume
one and the same deviation of the functions $F_1$ and $F_2$
from the mass-shell, and are of a higher order in $\gamma$ in 
the incoherent limit $\{F_1\}_{\rm corr}\, \{F_2\}_{\rm corr}$.
The decomposition rules above imply expansion in 
orders of $\gamma$ and differ from those used in ~\cite{SLM,SL}
where the decomposition is in the pole and off-pole terms.

\subsection{Examples}

Let us illustrate the use of correlation brackets
by writing the two-body $T$-matrix  and one-loop polarization 
in the next to leading order approximation. These functions, as well known,
contain the full information on the two-body scattering 
in the particle-particle and particle-hole channels respectively;
apart from the fact that they represent one of the few examples where the 
resummation series can be summed-up in a closed form,  their significance  
is due to the fact that the pole of the $T$-matrix signals
the onset of the superfluid phase transition (Cooper phenomenon), e.g. ~\cite{KB,DAN,THOULESS,SAL}, 
while that of the polarization function the onset 
of the growth of density fluctuations and 
liquid-gas phase transition, e.g. ~\cite{HEISELBERG,GRKMTA}. 

The contour ordered $T$-matrix equation reads:
\be
\label{TMAT}
{\bT}_{1234}= {\bV}_{1234} + i {\bV}_{1278}\otimes
{\bG}_{75}~ {\bG}_{86}\otimes {\bT}_{5634},
\ee
where  ${\bV}_{1234}= \bbox{\sigma}_z\, {V}_{1234}$, is the 
time-local bare interaction. In general, each element 
of the matrix ${\bT}$ is a function of four time arguments. 
The time locality of the potential, however,
implies a double-time structure  identical to eq. (\ref{2}). 
For the same reason, 
the particle-particle propagator product 
$\bG\,\bG \equiv {\bbox{\bG}_{[pp]}}$ should 
be considered as a single matrix. The components of the scattering
amplitudes, needed for complete specification of the self-energies, 
can be chosen as the retarded/advanced ones; the remaining 
components are provided by the optical theorem. 
For the retarded/advanced $T$-matrix
the Bethe-Salpeter equation in the Wigner (or mixed) 
representation reads
\be\label{TMIX}
\langle \bp\vert T^{R/A}(P, x)\vert \bp'\rangle &=&
\langle \bp \vert V\vert \bp'\rangle+i\sum_{{\bf p}''}  
\langle \bp \vert V\vert \bp''\rangle 
G_{[pp]}^{R/A}(\bp'', P, x)
\langle \bp''\vert T^{R/A}(P, x)\vert \bp' \rangle ,
\ee
where the leading order terms in the gradient expansion 
of the product ${\cal G}^{R/A} \, T^{R/A}$ have been kept. 
Here the subscript $[pp]$ indicates the particle-particle 
channel and $\bp$, $P$ relative momentum and total 
four-momentum respectively. The two-particle Green's function 
appearing in the kernel of equation (\ref{TMIX}) is given by
\be \label{G2}
G_{[pp]}^{R/A}(\bp_1, P_1, x) &=& \sum_{\omega_1}
\sum_{P_2}
\Big\{G^{>}\left(P_2/2+p_1, x\right) \, 
G^{>}\left(P_2/2-p_1, x\right) \nonumber \\
&-&G^{<}\left(P_2/2+p_1, x\right)\, G^{<}\left(P_2/2-p_1,x\right)\Big\}
\, {R}^{R/A} (P_1, P_2, x),
\ee
where 
\be\label{G20}
{R}^{R/A} (P_1, P_2,x) &=& 
\frac{(2\pi)^3\, \delta^3({\bP_1}-{\bP_2})}{E_1(x)-E_2(x)\pm i\eta}
\ee
is the free two-particle resolvent. The full 
off-mass-shell kernel now can be decomposed as 
\be\label{GEXP}
G_{[pp]}^{R/A}(\bp_1, P_1,x)&\simeq &
\Bigl[1-f(\ep_+,x)-f(\ep_-,x) \nonumber \\
&-&\{f(\ep_+,x)\}_{\rm corr} - \{f(\ep_-,x)\}_{\rm corr}\Bigr]\, 
{R}^{R/A}(E_1, \ep_+ + \ep_-,x) \nonumber \\
&+& \left[1-f(\ep_+,x)-f(\ep_-,x)\right]\, 
\Bigl[\{{R}^{R/A}(E_1,\ep_+,x)\}_{\rm corr} \nonumber \\
&+ & \{{R}^{R/A}(E_1,\ep_-,x)\}_{\rm corr}\Bigr]
\nonumber \\
&-&\left[ \{f(\ep_+,x)\, {\cal R}^{R/A}(E_1,\ep_+,x)\}_{\rm corr}
+ \{f(\ep_-,x)\, {\cal R}^{R/A}(E,\ep_-,x)\}_{\rm corr}\right] \nonumber\\
&+& {\cal O}\left(\gamma^2 \right),
\ee
where abbreviations $\ep_+\equiv \ep_{\bbox{P_1}/2+\bbox{p_1}}$ 
and $\ep_-\equiv \ep_{\bbox{P_1}/2-\bbox{p_1}}$ has been used; 
the relation between the propagators
$G^<,\, G^>$ and the distribution functions $f$ (i.e. the ansatz)
is given in the next subsection by equations (\ref{A1}) and (\ref{A2}).
The correlation brackets for the ${\cal R}$ function are more complicated 
than those defined above; here the convention is adopted to denote only 
the energy argument in the sum $\ep_++\ep_-$ with respect to which the 
correlation is constructed, i.e.
\be 
\label{G20CORR}
\{{\cal R}^{R/A}_0(E,\ep_{+})\}_{\rm corr} &=& \frac{1}{2}\sum_{\omega}
\gamma (\omega)\frac{{\cal P'}}{\ep_{+} - \omega}
\left[ {\cal R}^{R/A}_0(E,\omega+\ep_-)-{\cal R}^{R/A}_0(E,\ep_+ +\ep_-)\right],\\
\label{NGCORR}
\{f(\ep_+)\, {\cal R}^R_0(E,\ep_+)\}_{\rm corr} &=&
\frac{1}{2}\sum_{\omega} 
\, \gamma (\omega)\frac{{\cal P'}}{\ep_+ - \omega}
\, \left[ f(\omega)-f(\ep_+) \right]\nonumber\\
&\times&\left[ {\cal R}^{R/A}_0(E,\omega+\ep_-)
-{\cal R}^{R/A}_0(E,\ep_+ +\ep_-)\right].
\ee

The first term (in brackets) 
in eq. (\ref{G20}) corresponds to two on-mass-shell 
propagating  quasiparticles with intermediate state phase space occupied by 
both the quasiparticles and off-shell  
excitation. If the correlation terms are neglected 
one recovers the quasiparticle limit of the Bethe-Salpeter equation.
Since in the equilibrium limit the correlated states obey
Bose statistics  one may conclude that, apart from  
the Pauli-blocking due to the quasiparticles, the intermediate 
state propagation is Bose-enhanced by the off-shell
excitations. In nonequilibrium situations, of course, 
the occupations of quasiparticle and correlated states are the solutions
of the respective kinetic equations 
(see  equations (\ref{KINEQ3}) and 
(\ref{KINEQ4}) below).
The second term in brackets
corresponds to a two-particle propagation, 
where one of these is in an off-shell state; consistent with the 
next-to-leading order expansion, the intermediate state propagation 
is suppressed by quasiparticle Pauli-blocking. 
To the next-to-leading order in $\gamma$ the third term 
 vanishes in the incoherent limit; 
 in the coherent limit only an
 equilibrium treatment is possible. 
The kinetic theory, in fact, does not provide an
equation determining the time evolution of the off-mass-shell 
distribution function.
It is worthwhile to note that the  hole-hole 
propagation (the $G^{>}G^{>}$ term)  
is included in the two-particle propagator. When this term is 
dropped the $T$-matrix equation in the zero temperature 
limit reduces to a Brueckner-Galitskii~\cite{BRUECKNER,GALITSKI} 
type integral equation for a slightly non-perfect Fermi-gas.

The decomposition for the one-loop polarization function requires
a decomposition of the intermediate state particle-hole propagator 
\be \label{G22}
G_{[ph]}^{R/A}(\bp_1, P_1, x) &=&  \sum_{\omega_1}
\sum_{P_2}
\Big\{G^{>}\left(P_2/2+p_1, x\right) \, 
G^{<}\left(P_2/2-p_1, x\right) \nonumber \\
&-&G^{<}\left(P_2/2+p_1, x\right)\, G^{>}\left(P_2/2-p_1,x\right)\Big\}
\, {R}^{R/A} (P_1, P_2, x) .
\ee
Since one can trace a complete analogy to the $T$-matrix case we shall skip further details.

\subsection{The ansatz}
The kinetic equation (\ref{KB}) is still incomplete;
it should be supplemented  by a  relation between 
the functions $G^<$ and $G^>$. 
A natural choice is the Kadanoff-Baym ansatz, 
\be \label{A1}
-iG^{<}(p, x) &=& a(p, x) \, f(p, x),\\ \label{A2} 
iG^{>}(p, x) &=& a(p, x) \,\left[1- f(p, x)\right]. 
\ee 
The spectral function has already been defined as $a(p,x) 
= i\left[G^R(p,x) - G^A(p,x)\right]
\equiv i\left[G^>(p,x) - G^<(p,x)\right]$, therefore the 
ansatz replaces 
one of the propagators by 
the function $f(p, x)$ which  has the meaning of a 
quasiparticle distribution function; it reduces to 
the Fermi distribution in the equilibrium limit. The non-diagonal 
elements of the self-energy matrix can be expressed, similarly,  via
the spectral width and a distribution function,
\be\label{SIGGAM}\label{S1}
i\Sigma^{<}(p, x) &=& \gamma(p, x) \, f'(p, x),\\ \label{S2} 
-i\Sigma^{>}(p, x) &=& \gamma(p, x) \,\left[1- f'(p, x)\right].
\ee 
Though generally $f'(p, x) \not= f(p,x)$,  these functions, however, 
differ by terms of higher order in the gradient expansion
than is need for the present discussion (see \cite{BMALF}).
The reduction of the drift term in equation (\ref{KB})
can be facilitated by introducing auxiliary Green's functions,
\be \label{G0}
-i G^<_{\gamma\to 0} (p,x) &=&  2\pi\,  f(p, x)\, \delta(\omega -\ep_p(x)),
\quad
G^<_{\rm ren} (p,x) = [1-z(\bp, x)]\, G^<_{\gamma\to 0} (p,x)\\\label{G01}
 G^<_{\gamma} (p,x) &=&  -G^<_{\rm ren} (p,x)
+\Sigma^<(p,x) \frac{{\cal P'}}{\omega -\ep_p(x)},  
\ee 
and similarly for the $G^>(p,x)$ function with particle 
occupation replaced by the hole one,  $f(p) \to [1-f(p)]$. 
The  measure of the change of  quasiparticle Green's function due to 
the  wave-function renormalization is included 
in the $G_{\rm ren}^<$ propagator, where ``ren'' stands for 
renormalization.
In terms of these functions the expansion of
off-diagonal Green's functions with respect 
to the width  $\gamma$ is
\be\label{GSUM} 
 G^{<} (p,x) = G^{<}_{\gamma\to 0}(p,x)+G^{<} _{\gamma}(p,x).
\ee
If this decomposition is substituted  in the 
kinetic equation (\ref{KB}), the drift terms corresponding to the 
leading and next-to-leading order contributions  decouple:
\be\label{KING0} 
i\left\{{\rm Re}\, G^{-1}, \, 
G^{<}_{\gamma\to 0}-G^<_{\rm ren}\right\}_{\rm P.B.} 
&=& \Sigma^{>} G^{<}-\Sigma^{<}G^{>} , \\\label{KING1}
\left\{{\rm Re}\, G^{-1}, \, G^{<}_{\gamma} +G^<_{\rm ren}
\right\}_{\rm P.B.} & =& - \left\{\Sigma^{<},{\rm Re}\, 
G\right\}_{\rm \ P.B.}.
\ee
The decoupling of eqs. (\ref{KING0}) and (\ref{KING1}) derived in a slightly
different manner in ref. ~\cite{SL}, justifies the  
common practice of dropping the term $\left\{\Sigma^{<},{\rm Re}\, 
G\right\}_{\rm \ P.B.}$ from the 
Kadanoff-Baym equations when taking the QP limit.

\subsection{Single time equations}
Now we are in a position to derive the 
single time kinetic equations corresponding 
to  equations (\ref{KING0}) and (\ref{KING1}). Integrating these equations
over $\omega$ one finds
\be\label{KINEQ3} 
\left\{\frac{\partial}{\partial t}+\frac{\partial\ep_p}{\partial\bp} 
~\frac{\partial}{\partial\br} 
 - \frac{\partial\ep_p}{\partial \br}~ 
\frac{\partial}{\partial \bp}\right\}~f(\bp, \br, t) &=& I(\bp, \br, t),
\\\label{KINEQ4} 
\left(\frac{\partial}{\partial t}+\frac{\bp}{m}\frac{\partial}{\partial\br}\right) 
\{f(\bp,\br,t)\}_{\rm corr} 
&=&\frac{\partial }{\partial t} ~I_t (\bp,\br,t)
+\frac{\partial }{\partial \br}~ I_r(\bp,\br,t),
\ee
where the  collision integrals are
\be 
I(\bp,\br,t)&=&\int\frac{d\omega}{2\pi}
\left[\Sigma^{>}(\omega,\bp, \br, t)~G^{<}(\omega,\bp, \br, t)
-\Sigma^{<}(\omega,\bp, \br, t)~G^{>}(\omega,\bp, 
\br, t)\right],\\
I_t(\bp,\br,t)&=&\int\!\frac{d\omega\, d\omega'}{(2\pi)^2}
\, \frac{\cal P'}{\omega-\omega'}\, 
\left[\Sigma^{>}(\omega,\bp,\br, t)\, G^{<}(\omega',\bp,\br, t)
- \Sigma^{<} (\omega',\bp,\br, t)\, G^{>}(\omega,\bp,\br, t)\right], \\
I_r(\bp,\br,t) &=&\int\!\frac{d\omega\, d\omega'}{(2\pi)^2}
\frac{{\cal P}}{\omega-\omega'}\,  
\Biggl\{ G^{<}(\omega,\bp,\br, t)\, \frac{\partial}{\partial \bp} 
\Bigl[ \Sigma^{<}(\omega',\bp,\br, t)-\Sigma^{>}(\omega',\bp,\br, t) \Bigr]
\nonumber\\
&+&\Sigma^{<}(\omega, \bp,\br, t)\, \frac{\partial}{\partial \bp} 
\Bigl[G^{>}(\omega',\bp, \br, t)-G^{<}(\omega',\bp,\br, t)\Bigr]\Biggr\}.
\ee
Equations (\ref{KINEQ3})  and (\ref{KINEQ4}) couple the time evolutions 
of the distribution functions for the on-mass-shell propagating 
quasiparticles and off-shell excitations, 
which are described by the distribution functions $f(\bp,\br,t)$, and
$\{f(\bp,\br,t)\}_{\rm corr}$ respectively. The terms involving 
space-time derivatives in (\ref{KINEQ4}) on the right- and left-hand 
side of the equations balance each other separately; e.g. the correlated 
distribution function can be determied from $I_t$ by a direct 
integration (provided the initial conditions are known). 
The latter collision integral involves only the on-mass-shell
singlet distribution functions, whose time-evolution is governed by the 
kinetic equation (\ref{KINEQ3}). One may now compare the 
kinetic equations (\ref{KINEQ3}) and (\ref{KINEQ4}) to those of refs. 
~\cite{KMGR,SLM,SL}. Ref. ~\cite{KMGR} gives a single
kinetic equation for the Wigner distribution function. Summing eqs. 
(\ref{KINEQ3}) and (\ref{KINEQ4}) we find a kinetic equation which is 
consistent with the eq. (49) of ref. ~\cite{KMGR}, except the terms involving 
the space gradients of  the correlated distribution function. This difference
lies presumably in a different partition of the collision and mean-field terms
(note that the both terms are governed by the same timescales). 
Present form seems to be more suited for establishing the conservation laws;
e.g. the number density conservation is obtained automatically by taking the 
momentum integral, while ref. ~\cite{KMGR} drops the space gradients. 
Refs. ~\cite{SLM,SL} present two kinetic equations; compared to their 
equation for the  singlet distribution function, which is 
defined as the pole values of the Wigner function, our eq. (\ref{KINEQ3}) differs
by the fact that it includes  only the $\gamma \to 0$ part of the
Wigner function. The second kinetic equation ref. ~\cite{SL} gives in
the integral form, which is evaluated in the Born approximation. 
The content of our equation (\ref{KINEQ4}) is again different, 
since it includes the dynamics of all correlations $\propto \gamma$. 
The collision integrals, as far as we can judge, differ considerably
in their general form  (cf. eq. (56) in ref. ~\cite{SL}).

Summing  eqs. (\ref{KINEQ3}) and (\ref{KINEQ4}) and integrating over momentum 
we find the particle number conservation 
\be\label{DC} 
\frac{\partial}{\partial t}\left[
n_{\rm free}(\br, t)+n_{\rm corr}(\br , t)
\right] 
+ {\bf  \nabla}_{\br} \left[\bj_{\rm free}(\br, t)
+\bj_{\rm corr}(\br , t)\right] = 0  ,
\ee
where 
\be 
n_{\rm free}(\br, t)& =& \int\!\frac{d^3p}{(2\pi)^3} \, f(\bp, \br, t),\\
n_{\rm corr}(\br, t) &=& \int\!\frac{d^3p}{(2\pi)^3} \, \{f(\bp, \br, t)\}_{\rm corr},\\
\bj_{\rm free}(\br, t) &=& \int\!\frac{d^3p}{(2\pi)^3}\,  
\frac{\partial\ep_p}{\partial \bp} \,  f(\bp, \br, t),\\
\bj_{\rm corr}(\br , t)&=&  \int\!\frac{d^3p}{(2\pi)^3}\, \frac{\partial
  \ep_p}{\partial \bp} \, \{f(\bp, \br, t)\}_{\rm corr},
\ee
and the terms ${\cal O} (\gamma^2)$ have been 
dropped in the expression for the current of off-shell excitations, 
$\bj_{\rm corr}(\br , t)$; (symmetrization of the  collision integrals in the 
usual manner shows that they vanish after integration over momentum). 
The separation of the Wigner distribution function
$f_{\rm tot} (\bp,\br,t)$ in leading and next-to-leading order terms, as noted above,
is not the only possible one. In fact, the separation in the pole and off-pole 
terms has been employed~\cite{SL}, where 
\be\label{DENSMAT}
f_{\rm tot}(\bp) = \int\!\frac{d\omega}{2\pi i}\,
\left[G^{<}_{\gamma\to 0}(\omega,\bp)+G^{<}_{\gamma}
(\omega,\bp)\right] =  z(\bp)\,  f(\bp) +  f_{\rm off}(\bp)
\ee
with
\be 
f_{\rm off}(\bp,\br,t) = \int \frac{d\omega}{2\pi} 
[G^{<}_{\gamma}(\omega,\bp)-G^{<}_{\rm ren}(\omega,\bp)].
\ee
Though formally equivalent, the latter partition  does not fulfill the 
frequency sum rule at each order of the expansion and one has to
sum up at the least  first two terms.  The partition in powers of $\gamma$
has the advantage that it  does so at any order of the expansion and, 
in addition, it permits a simple interpretation of various terms 
because, close to equilibrium, the distribution function of correlated 
exciatations have a Bose-type spectrum.

\section{The Three-Body Problem}

We proceed to formulate the three-body problem at finite temperatures and 
densities.
The resummation series for three-body scattering amplitudes have the general form,  
\be\label{TFULL}
\hat {\bcT}&=& \hat {\bcV} + \hat {\bcV}\otimes \hat {\bcG}\otimes\hat {\bcV}
\nonumber \\
&=& \hat {\bcV} + \hat {\bcV}\otimes\hat {\bcG}_0\otimes\hat{\bcT}
= \hat {\bcV} + \hat{\bcT}\otimes\hat {\bcG}_0\otimes\hat {\bcV}
\ee  
where  $\hat {\bcG}_0$  and $\hat {\bcG}$ are the free and full three-particle 
Green's functions,  and $\hat{\bcV}$ is the interaction (we use the operator form 
for notational simplicity; each operator, as in the 
two-particle case, is combined in a $2\times 2$ matrix, with 
elements defined on the contour).  In the case of pair-interactons, 
the net interaction is  $\hat {\bcV} = \hat {\bcV}_{12}+\hat {\bcV}_{23}
+\hat {\bcV}_{13}$, where $\hat{\bcV}_{\alpha\beta}$ is the 
interaction potential between 
particles $\alpha$ and $\beta$; note that  the potential 
matrices are time diagonal due to the time-locality of the  potential. The kernel of  eq. (\ref{TFULL}) 
is not square integrable: the pair-potentials
introduce delta-functions due to momentum conservation for the spectator 
non-interacting particle and the iteration series contain  singular terms 
(e.g., of type $\hat {\bcV}_{\alpha\beta}\hat{\bcG}_0 
\hat {\bcV}_{\alpha\beta}$ to  the lowest order in the interaction). 
The resulting ambiguity is eliminated by a rearrangment, 
originally due to Faddeev ~\cite{FADDEEV}, 
which plugs the singular  term in the channel $\hat {\bcT}_{\alpha\beta}$-matrices 
corresponding to the case when the interaction between the particles 
$\alpha ,\beta$ and the third particle $\gamma$
is neglected. The total $\hat {\bcT}$-matrix 
has a decomposition 
\be \label{DECOMP}
\hat {\bcT} = \hat {\bcT}^{(1)}+ \hat {\bcT}^{(2)} + \hat {\bcT}^{(3)}, 
\ee
where 
\be\label{TUP}
\hat {\bcT}^{(\alpha)} &=& \hat {\bcV}_{\beta\gamma} + 
\hat {\bcV}_{\beta\gamma}\otimes \hat {\bcG}_0\otimes \hat {\bcT}.
\ee
and   $\alpha\beta\gamma=123,\, 231,\, 312$.  
The  channel $(\alpha\beta)$ transition operators
resum the  successive iterations with driving term $\hat {\bcV}_{\alpha\beta}$,
\be\label{TDOWN}
\hat {\bcT}_{\alpha\beta}=  \hat {\bcV}_{\alpha\beta} + \hat
 {\bcV}_{\alpha\beta}\otimes \hat {\bcG}_0\otimes \hat {\bcT}_{\alpha\beta};
\ee
they are directly related to the two-body effective interaction in the 
$\alpha\beta$-channel, e. g. two particle $T$-matrix, eq. (\ref{TMAT}).
In terms of these functions the total three-body amplitude is completely determined 
via the three coupled integral equations 
\be\label{FTFA}
\hat{\bcT}^{(\alpha)}=\hat  {\bcT}_{\beta\gamma} +
\hat{\bcT}_{\beta\gamma}\otimes \hat{ \bcG}_0\otimes\left(
\hat{\bcT}^{(\beta)}+\hat{ \bcT}^{(\gamma)}\right),
\ee
which represent non-singular  Fredholm-type integral equations; their
formal structure is identical to the Faddeev equations in the  vacuum~\cite{FADDEEV}. 
To see the statistical nature of the system (finite temperature 
and density) which enters via the contour ordering of the operators reflected
in their matrix structure, let us write eq. (\ref{FTFA})
in the momentum representation. E. g. for   the retarded component of 
$\hat {\bcT^{(1)}}$ one finds
\be\label{FTFA0} 
\langle \bk_{23} ,\bq_1\vert {\cal T}^{R\,(1)}(t, t',\bK) \vert \bk_{23} ' ,\bq_1 ' \rangle
&=& \langle \bk_{23} ,\bq_1\vert {\cal T}^{R}_{23}(t, t',\bK) \vert \bk_{23} ' ,\bq_1 '\rangle
\nonumber \\
&+&\int \Biggl\{ \Big\langle \frac{\bk_{23}}{2}-\frac{3 \bq_1}{4}, -\bk_{23} -\frac{\bq_1}{2}
\Big\vert {\cal T}^{R\, (2)}(t, t''',\bK) \vert \bk_{13}''' ,\bq_2'''\rangle \nonumber \\
&+& \Big \langle 
\frac{\bk_{23}}{2}+\frac{3\bq_1}{4}, -\bk_{23} +\frac{\bq_1}{2} 
\Big \vert {\cal  T}^{R\, (3)}(t, t''',\bK) \vert \bk_{13}''' ,\bq_{2}'''
\rangle\Biggr\}\nonumber \\
&\times&\langle \bk_{13}''',\bq_{2}'''\vert {\cal G}_0^{R}(t''', t'',\bK) 
\vert \bk''_{13},\bq_{2}''\rangle\nonumber \\
&\times&
\Big\langle\frac{\bk''_{13}}{2} -\frac{3\bq''_{2}}{4}, -\bk''_{13} 
-\frac{\bq_{2}''}{3}\Big \vert {\cal T}^{R}_{23}(t'', t',\bK) \vert\bk_{23}' ,\bq_1'
\rangle \nonumber \\
&\times& \bk_{13}''' d\bq_{2}'''dt''' d\bk_{13}'' d\bq_{2}''  dt'',
\ee
where we used the time-locality of the interaction;
(the obvious dependence of the functions on $x$ has been dropped).
Here the momentum space is spanned in terms of Jacobi coordinates,
\be\label{JACOBI}
\bK = \bp_{\alpha}+\bp_{\beta}+\bp_{\gamma},\quad \bk_{\alpha\beta} = \frac{1}{2}(\bp_{\alpha} - \bp_{\beta}),\quad
\bq_{\gamma} = \frac{1}{3} (\bp_{\alpha}+\bp_{\beta}) 
- \frac{2}{3}\bp_{\gamma}.
\ee
The free three-particle Green's function has different types of factorizations
depending on the particle-hole content of the three-body $\cal T$-matrix. 
One may identify four types of direct factorizations
\be 
{\cal G}_{0\left[ppp\right]}^R(t_1,t_2) &=& 
\theta (t_1-t_2)\,\left[G^{>}(t_1,t_2)G^{>}(t_1,t_2) 
G^{>}(t_1,t_2)-G^{<}(t_1,t_2)G^{<}(t_1,t_2)G^{<}(t_1,t_2)\right], \\
%%%%%%%%%%
{\cal G}_{0\left[pph\right]}^R(t_1,t_2) &=& 
\theta (t_1-t_2)\,\left[G^{>}(t_1,t_2)G^{>}(t_1,t_2) 
G^{<}(t_1,t_2)-G^{<}(t_1,t_2)G^{>}(t_1,t_2)G^{<}(t_1,t_2)\right] ,\\
%%%%%%%%%%
{\cal G}_{0\left[phh\right]}^R(t_1,t_2) &=& 
\theta (t_1-t_2)\,\left[G^{>}(t_1,t_2)G^{<}(t_1,t_2) 
G^{<}(t_1,t_2)-G^{<}(t_1,t_2)G^{>}(t_1,t_2)G^{>}(t_1,t_2)\right], \\
%%%%%%%%%%%%
{\cal G}_{0\left[hhh\right]}^R(t_1,t_2) &=& 
\theta (t_1-t_2)\,\left[G^{<}(t_1,t_2)G^{<}(t_1,t_2) 
G^{<}(t_1,t_2)-G^{>}(t_1,t_2)G^{>}(t_1,t_2)G^{>}(t_1,t_2)\right],
\ee 
where subscripts $[p..]$ and $[h..]$ refer to particle and hole states
respectively. The first line corresponds
 the 3-particle channel which is dominant in Fermi-systems 
interacting via short-range forces, and to which we shall restrict ourselves.
The Faddeev decompositon, eq. (\ref{FTFA0}), and
five exchange diagrams for this case are shown in Fig. 1. 

The signs of the exchange diagrams are given by $(-1)^n$, where $n$ is 
the number of the line intersections in the diagram. 
The respective exchange terms for the remainder of the  diagrams are obtained 
by interchanging the time-direction of lines.

A Fourier transformation in eq. (\ref{FTFA0}) with respect to the relative 
times leads to the 3-particle ${\cal T}$-matrix 
 \be\label{FTFA1} 
\langle \bk_{23} ,\bq_1\vert {\cal T}^{R\,(1)}(K) \vert \bk_{23} ' ,\bq_1 ' \rangle
& =& \langle \bk_{23} ,\bq_1\vert {\cal T}^{R}_{23}(K) 
\vert \bk_{23} ' ,\bq_1 '\rangle\nonumber \\
&+&\int \Bigl\{ \Big \langle \frac{\bk_{23}}{2}-\frac{3 \bq_1}{4}, -\bk_{23} -\frac{\bq_1}{2}
\Big \vert {\cal  T}^{R\, (2)}(K) \vert \bk_{13}'' ,
\bq_2''\rangle\nonumber \\
&+& \Big \langle \frac{\bk_{23}}{2}+\frac{3\bq_1}{4}, 
-\bk_{23} +\frac{\bq_1}{2} \Big \vert {\cal T}^{R\, (3)}( K) \vert \bk_{13}'' ,\bq_{2}''\rangle\Bigl\}\nonumber \\
&\times&
\frac{Q(K', k_{13}'', q_2'')\delta(\bK-\bK')}{\Omega -\Omega'+i\eta}
\nonumber \\
&\times&
 \Big\langle\frac{\bk''_{13}}{2} -\frac{3\bq''_{2}}{4}, -\bk''_{13} 
-\frac{\bq_{2}''}{3}\Big \vert {\cal T}^{R}_{23}( K) \vert\bk_{23}' ,\bq_1'
\rangle \nonumber \\
&\times&\, d\omega_{13}'' d\bk_{13}'' 
d \nu_2'' d\bq_{2}'' d\Omega'   d\bK' ,
\ee
with 
\be 
Q(K,k_{13}, q_2) &=& \Bigl\{
G^{>}\left(K/3+k_{13}+q_2/2\right)
G^{>}\left(K/3-k_{13}+q_2/2\right)
G^{>}\left(K/3-q_2\right)\nonumber \\
&-&G^{<}\left(K/3+k_{13}+q_2/2\right)
G^{<}\left(K/3-k_{13}+q_2/2\right)
G^{<}\left(K/3-q_2\right)\Bigr\},
\ee
and four-vector notation  $K=(\Omega, \bK)$, $k =(\omega, \bk)$, 
$q=( \nu, \bq)$. The momentum representation of equations for 
${\cal T}^{(2)}$ and ${\cal T}^{(3)}$ follows  from (\ref{FTFA1}) by permutations of (Greek) indexes $1\leftrightarrow 2$ and  $1\leftrightarrow 3$. 

\begin{figure}
\begin{center}
\mbox{\psfig{figure=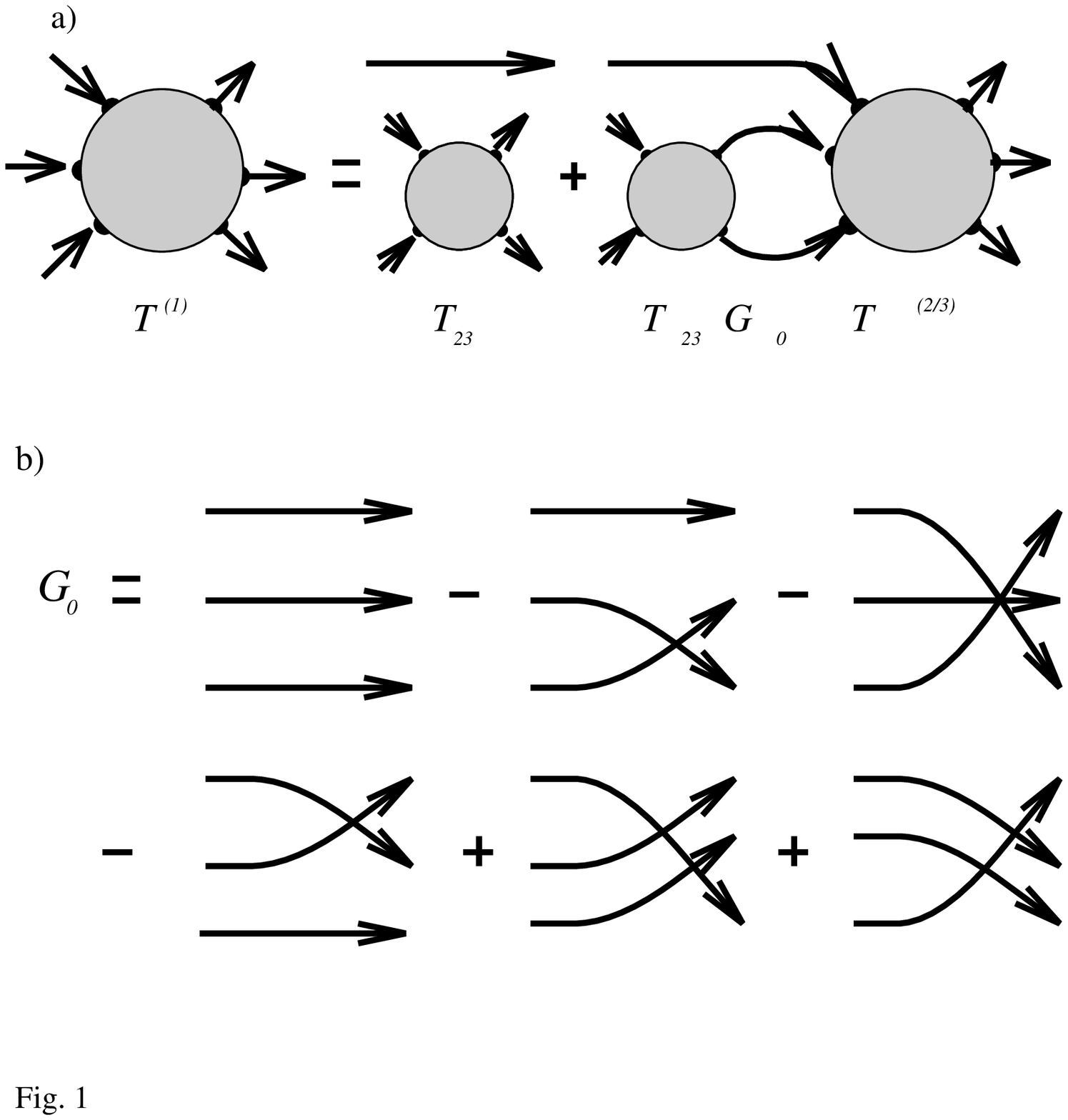,height=2.75in,width=3.4in,angle=0}}
\end{center}
\caption[] 
{\footnotesize{a) The three-body amplitudes in the 3-particle 
channel after Faddeev rearrangement; only 
one of the terms of the type ${\cal T}^{(\alpha)}$ appearing on the right hand side is shown. 
b) The direct and five exchange three particle propagators. The 
diagrams for processes involving hole 
propagations are obtained by interchanging the direction of the lines. 
}}
\label{fig1}
\end{figure}
To the lowest order in the expansion of the off-diagonal single-particle 
Green's function we find the QP result   
 \be\label{FTFA2} 
\langle \bk_{23} ,\bq_1\vert {\cal T}^{R\,(1)}(K) \vert 
\bk_{23} ' ,\bq_1 ' \rangle
& =& \langle \bk_{23} ,\bq_1\vert {\cal T}^{R}_{23}(K) 
\vert \bk_{23} ' ,\bq_1 '\rangle\nonumber \\
&+&
\int \Bigl\{ \Big \langle \frac{\bk_{23}}{2}-\frac{3 \bq_1}{4}, -\bk_{23} -\frac{\bq_1}{2}
\Big \vert  {\cal T}^{R\, (2)}(K) \vert \bk_{13}'' ,\bq_2''\rangle \nonumber \\
&+& \Big \langle \frac{\bk_{23}}{2}+\frac{3\bq_1}{4}, -\bk_{23} 
+\frac{\bq_1}{2} 
\Big \vert {\cal T}^{R\, (3)}(K) \vert \bk_{13}'' ,\bq_{2}''\rangle\Bigl\}
\, Q(\bK, \bk_{13}'', \bq_2'')
\nonumber \\&\times&
\frac{ \Big\langle\frac{\bk''_{13}}{2} 
-\frac{3\bq''_{2}}{4}, -\bk''_{13} -\frac{\bq_{2}''}{3}
\Big \vert {\cal T}^{R}_{23}
( K) \vert\bq_{23}' ,\bq_1'\rangle}
{\Omega-\epsilon_{\vec K/3+\vec k_{13}''+\vec q_2''/2}-\epsilon_{\vec
K/3-\vec k_{13}''+\vec q_2''/2}-\epsilon_{\vec K/3-\vec q_2''}+i\eta} 
   d\bk_{13}''  d\bq_{2}''  ,\nonumber \\
\ee
with 
\be\label{Q}
Q(\bK, \bk_{13}, \bq_2)&=& 1-f\left(\bK/3+\bk_{13}+\bq_2/2\right) 
-f\left(\bK/3-\bk_{13}+\bq_2/2\right)-f\left(\bK/3-\bq_2\right)\nonumber \\
&-& f\left(\bK/3+\bk_{13}+\bq_2/2\right)
f\left(\bK/3-\bk_{13}+\bq_2/2\right)\nonumber \\
&-&f\left(\bK/3+\bk_{13}+\bq_2/2\right)f\left(\bK/3-\bq_{2}\right) \nonumber \\
&-& f\left(\bK/3-\bk_{13}+\bq_2/2\right)f\left(\bK/3-\bq_2\right).
\ee
Equations (\ref{FTFA1}) and (\ref{FTFA2}) provide
the Faddeev amplitudes appropriate at 
finite densities and temperatures. The medium effects enter 
these equations in two ways: 
(i) the renormalization of the single particle spectrum 
in the resolvent of eq. (\ref{FTFA1}),
and (ii) statistically occupied  intermediate state propagation.  
The first line of  eq. (\ref{Q}) is the Pauli-blocking of intermediate  
propagation of three particles and is the 
dominant term in the dilute limit. The 
remainder of the terms, in the same limit, 
tend to Bose-enhance the intermediate propagation,
which can be seen from  the equilibrium identity $f(\omega_1) f(\omega_2) = 
g(\omega_1+\omega_2)  \left[1-f(\omega_1)-f (\omega_2)\right]$, where $f$
and $g$ are the Fermi and Bose distribution functions. 
Inclusion of the next to leading order terms in the powers of $\gamma$ 
can be constructed in a manner similar to the two-body case, however would 
lead to implicite inclusion of four body correlations, which are beyond of 
the scope of this paper. In the limit $Q\to 1 $ these equations reduce to the original 
Faddeev equations (of course with the particle spectrum $\epsilon = p^2/2m$). 

\section{Scattering Integrals}

\subsection{Transport Vertex}
The self-energies in the matrix formulation of nonequilibrium 
theory have a structure topologically identical to the equilibrium case. 
We shall adopt the cluster expansion for the self-energy
matrix ~\cite{ROEPKE,KREMP}  and, 
consistent with the discussion above, truncate it at the three-body
level:
\be\label{SIGMAT} 
{\bf \Sigma}_{12} = {\bf T}_{1234} \otimes {\bf G}_{43^+} + 
{\bcT}_{165432}\otimes {\bf G}_{43^+}\,{\bf G}_{65^+},
\ee
where $+$ means an infinitesimal later time point.
The non-diagonal elements of the matrix equation provide the collision rates 
on the right hand side of the kinetic equations. In the mixed representation 
one finds 
\be\label{SIGMIN} 
\Sigma^{<}(p_1) &=& -i \sum_{p_2}\langle {(\bp_1 - \bp_2)/}{2}\vert 
 T^{<}(p_1+p_2,)\vert {(\bp_1 - \bp_2)/}{2}\rangle
\, 
G^{>}(p_2, )\nonumber\\
&+& \sum_{p_2,p_3}\Bigl[\langle \bk_{23}, \bq_1 \vert 
{\cal T}^{(1)\, <}(K)\vert \bk_{23}, \bq_1\rangle\,
+\langle \bk_{13}, \bq_2 \vert 
{\cal T}^{(2)\, <}(K)\vert \bk_{13}, \bq_2\rangle\,\nonumber\\
&+&\langle \bk_{12}, \bq_3 \vert  
{\cal T}^{(3)\, <}(K)\vert \bk_{12}, \bq_3\rangle\,
\Bigr] G^{>}(p_2)\,G^{>}(p_3).
\ee
The expression for $\Sigma^>$ follows by simultaneous interchange of
$<$ and $>$ signs. The Jacobi momenta in the  three-body ${\cal T}$-matrix
are related to the arguments of the self-energy and the Green's functions via 
the relations (\ref{JACOBI}). Equation (\ref{SIGMIN}) 
does not imply the usual form of the transition probability as a square 
of the transition amplitude times the statistical factors (i.e.
Fermi Golden Rule). For the purpose of recovering this form
one may use the non-equilibrium optical theorem
\be\label{TMIN}
\langle\bp_1 \vert T^{<}(P )\vert\bp_2\rangle &=& -i\sum_{p_3\, p_4}
\langle\bp_1\vert T^{R}(P)\vert 
\frac{\bp_3-\bp_4}{2}\Big\rangle~G^{<}(p_3 )\, G^{<}(p_4)
\Big\langle \frac{\bp_3-\bp_4}{2}\vert T^{A}(P)\vert\bp_2 
\rangle \nonumber \\
&\times &(2\pi)^4\, \delta^4(P-p_3-p_4) ,
\ee
\be\label{TMIN3}
\langle \bk_{23}, \bq_1\vert 
{\cal T}^{<}(K)\vert \bk_{23}', \bq_1'\rangle
&=& \sum_{\alpha,\beta}\sum_{p_4,p_5,p_6}\langle \bk_{23}, \bq_1\vert 
{\cal T}^{(\alpha)\, R}(K)\vert \bk_{45}, \bq_6\rangle\,
G^{<}(p_4)\,G^{<}(p_5)\,G^{<}(p_6)\nonumber \\
&\times& \langle \bk_{45}, \bq_6 \vert 
{\cal T}^{(\beta)\, A}(K)\vert \bk_{23}', \bq_1'\rangle 
(2\pi)^4\, \delta^4(K-p_4-p_5-p_6) .
\ee 
Combination of the equations 
(\ref{SIGMIN}),(\ref{TMIN}), and (\ref{TMIN3}) yields 
the usual form of the quantum Boltzmann collision integrals 
with scattering probabilities defined as 
\be 
\langle \bp_1 ,\bp_2 \vert W \vert \bp_3 , \bp_4\rangle 
 &=& \langle(\bp_1-\bp_2)/2\vert T^{R}(P)\vert 
{(\bp_3-\bp_4)/}{2}\rangle\langle 
{(\bp_3-\bp_4)}/{2}\vert T^{A}(P)\vert (\bp_1-\bp_2)/2 \rangle
\nonumber \\
&\times& \delta^3(\bp_1+\bp_2-\bp_3-\bp_4) \\ 
\langle \bp_1 ,\bp_2 ,\bp_3\vert {\cal  W }\vert \bp_4 , \bp_5 , \bp_6
\rangle &=& \sum_{\alpha,\beta} \langle \bk_{23}, \bq_1\vert
{\cal T}^{(\alpha)\, R}(K)\vert \bk_{45}, \bq_6\rangle\,
\langle \bk_{45}, \bq_6 \vert
{\cal T}^{(\beta)\, A}(K)\vert \bk_{23}, \bq_1\rangle \nonumber \\
&\times& (2\pi)^3\, \delta^3(\bp_1+\bp_2+\bp_3-\bp_4-\bp_5-\bp_6) , 
\ee
where it is understood that the Jacobi momenta  in the three-body
transition probability $\cal W$ should be expressed through the momenta 
$p_i$, $i = 1...6$, via the relations (\ref{JACOBI}).

The diagonal elements of the matrix equation (\ref{SIGMAT}) provide the 
solution of the integral Dyson equation and thus, the  
renormalization of the single particle spectrum. Apart from the details
(which can be deduced along the lines of the previous discussion, see
Section 1 E), 
let us note that (i) in the next-to-leading order approximation 
to the spectral function the mean-field due to the excitations 
occupying  off-mass-shell states contributes
to the the real part of the self-energy;
(ii)  an additional contribution to the self-energy
comes from the off-mass-shell $T$-matrices folded by the on-shell
quasiparticle distribution - a contribution which is 
missing in the usual Brueckner theory; (iii) 
the three-body processes contribute to the energy shift in 
the single particle spectrum via the second term of the cluster 
expansion.

\subsection{Virial expansion of collision integrals}

The general form of the collision integral can be arranged in the form
of virial corrections in much the same manner as the diagrams of 
scattering theory (Section 1 C). In principle, in the 
next-to-leading order approximation this collision integral 
can be separated into  the quasiparticle and correlated contributions
in two different ways,  corresponding to the partition in the pole and 
off-pole terms and in the powers of $\gamma$. The latter partition  
is physically more appealing, since it involves the coupling 
between the kinetic equations for on-shell quasipaticles and 
off-shell excitations. This decomposition for the two-body 
scattering integral contains five terms
\be 
I(\bp,x)=I^{(0)}(\bp,x)+I^{(1)}(\bp,x)+I^{(2)}(\bp,x)
+I^{(3)}(\bp,x)+I^{(4)}(\bp, x);
\ee 
a similar decomposition emerges
for the three-body collision integral ${\cal I}$.
The first terms in the decomposition i.e. the scattering integrals
for the quasiparticles, 
\be 
 I^{(0)}(\bp_1) &=& \sum_{\bf p_2p_3p_4}
\langle \bp_1, \bp_2\vert W\vert\bp_3 , \bp_4\rangle
\delta(\ep_{p_1}+\ep_{p_2}-\ep_{p_3}-\ep_{p_4})\nonumber \\
 &\times &\left\{f(\bp_1)\, f(\bp_2) [1-f(\bp_3)] [1-f(\bp_4)] 
-[1-f(\bp_1)] [1-f(\bp_2)]f(\bp_3)\, f(\bp_4)\right\}, \\
{\cal I}^{(0)}(\bp_1) &=& \sum_{\bf p_2p_3p_4 p_5 p_6}
 \langle \bp_1, \bp_2 ,\bp_3\vert W\vert\bp_4 , \bp_5 ,\bp_6\rangle
\delta(\ep_{p_1}+\ep_{p_2}+\ep_{p_3}-\ep_{p_4}
-\ep_5-\ep_6)\nonumber \\
 &\times &\Big\{f(\bp_1)\, f(\bp_2) \,  f(\bp_3) [1-f(\bp_4)] [1-f(\bp_5)] 
[1-f(\bp_6)]\nonumber\\
&-&[1-f(\bp_1)][1-f(\bp_2)][1-f(\bp_3)] f(\bp_4)\, f(\bp_5) \,  f(\bp_6)\Big\}, 
\ee 
describe the scattering of two/three incoming uncorrelated quasiparticles 
into uncorrelated outgoing states. Note that
 the retarded amplitudes in the collision probabilities 
include, apart from the usual Pauli-blocking  due to the 
uncorrelated quasiparticles, the off-shell intermediate propagation
effects and the phase space occupation due to the 
off-shell excitations in the medium. 

The remaining collision integrals describe 
scattering processes where one of the incoming or outgoing 
excitations is in a correlated state. (The upper index 
labels the correlated particle). Note, however, 
that the processes involving correlated states implicitly 
contain at least an extra quasiparticle. 
The terms of higher than zeroth order in the three particle 
collision integrals, therefore, involve processes with four or 
larger number of quasiparticles and can be ignored.
As to the two-body case, the result for the 
second term  of the decomposition reads
\be\label{ID1}
I^{(1)}(\bp_1) &=&
\sum_{{\bf p_2  p_3 p_4}}
\langle \bp_1, \bp_2\vert W\vert\bp_3 , \bp_4\rangle
\Bigl[\{f(\ep_{p_1})\}_{\rm corr}
\delta(\ep_{p_1}+\ep_{p_2}-\ep_{p_3}-\ep_{p_4}) \nonumber\\ 
&+&f(\ep_{p_1})\{\delta(\ep_{p_1})\}_{\rm corr}
+\{f(\ep_{p_1})\, \delta(\ep_{p_1})\}_{\rm corr}  \Bigr]\nonumber\\ 
&\times&\Bigl\{f(\bp_2)\left[1-f(\bp_3)\right]\left[1-f(\bp_4)\right] +\left[1-f(\bp_2)\right]f(\bp_3)\, f(\bp_4)\Bigr\} ;  
\ee 
where the functions $\{\delta(\ep_{p_1})\}_{\rm corr}$
and $\{f(\ep_{p_1})\, \delta(\ep_{p_1})\}_{\rm corr}$ are defined in 
complete analogy to eqs. (\ref{G20CORR}) and (\ref{NGCORR}).
The explicit expression for $I^{(2)}$ follows from  $I^{(1)}$ by interchanging
$\bp_1 \leftrightarrow \bp_2$, while those for  $I^{(3)}$ and $I^{(4)}$ from simultaneous interchanges $\bp_4 \leftrightarrow \bp_2,\, \bp_3 
\leftrightarrow \bp_1$ and $\bp_4 \leftrightarrow \bp_1,\, \bp_3 
\leftrightarrow \bp_2$, respectively. 
Consistent with keeping only the first order terms in $\gamma$,
the scattering amplitudes in the integrals $I^{(i)}$, $i=1,2,3,4$,
are those including only the quasiparticle contributions.
The first terms in the bracket in eq. (\ref{ID1})
describes scattering of excitations, 
when one of these is in an  initial or final 
correlated state. This term describes a {real process} 
with energy conservation in the scattering event and 
irreversible deformation of the wave functions of the 
scattering particles. The second term corresponds to the 
{\it virtual scattering} of quasiparticles with explicit 
nonconservation of energy and reversible 
deformation of the wave functions;
the third term is of the coherent nature discussed above
(see Fig. 2 a).

Had we preferred the decomposition 
in the pole and off-pole terms, then  the 
zeroth order term would have acquired an additional factor which is 
a product of the wave-function renormalizations of all four quasiparticles.  
For the next-to-leading order virial correction one finds
\be\label{I1}
\tilde I^{(1)}(\bp_1) &=&
\sum_{\bf p_2  p_3 p_4} \langle \bp_1 , \bp_2 \vert W\vert
\bp_3 , \bp_4\rangle\frac{{\cal P'}}
{\ep_{p_1}+\ep_{p_2}-\ep_{p_3}-\ep_{p_4}}
\gamma(\ep_{p_3}+\ep_{p_4}-\ep_{p_2})\nonumber \\
&\times& \Big\{f(\ep_{p_3}+\ep_{p_4}-\ep_{p_2})\, f(\bp_2)\nonumber\\
&\times&\left[1-f(\bp_3)\right]\left[1-f(\bp_4)\right]
-\left[1-f(\ep_{p_3}+\ep_{p_4}-\ep_{p_2})\right] \nonumber \\
&\times& \left[1-f(\bp_2)\right]f(\bp_3)\, f(\bp_4)\Big\} .  
\ee 
The fact that this collision integral corresponds to an effective three-body
process can be visualized by inserting in the latter equation the 
explicit expression for the spectral width (in the quasiparticle 
approximation)

\begin{figure}
\begin{center}
\mbox{\psfig{figure=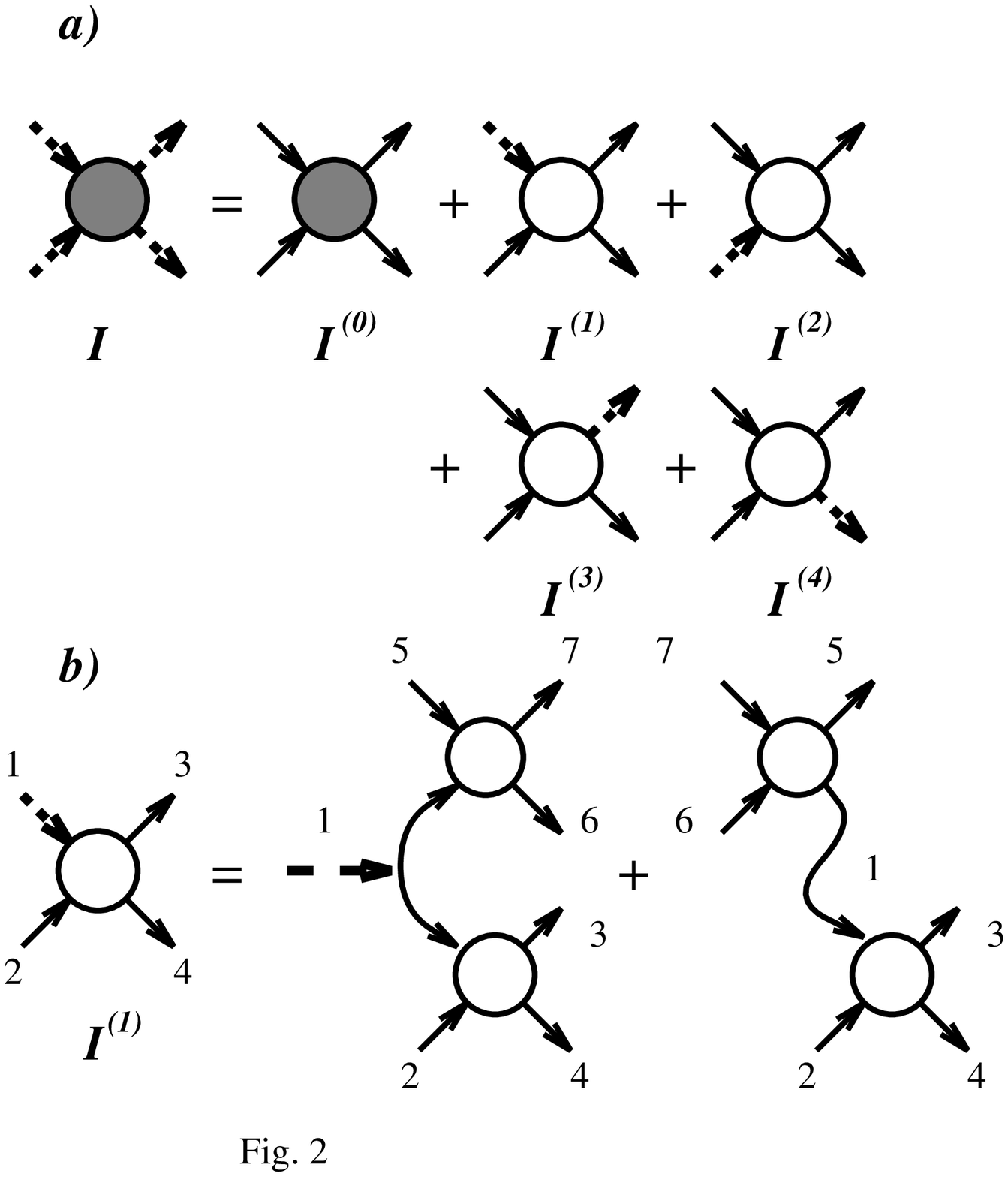,height=5.in,width=3.4in,angle=0}}
\end{center}
\caption[] 
{\footnotesize{ a)  The virial decomposition of the collision integral in the two
body case; the open circles denote the $T$ matrix in the quasiparticle 
approximation, while the filled ones include also the off-shell contributions.
b) The modification of next-to-leading order correction to the gain 
part of the collision integral
after substituting the explicit expression for the width; all the final state
particles are now on the mass-shell.}}
\label{f:gap}
\end{figure}

\begin{eqnarray}
  \label{GQPA}
\gamma (\bp_1) &=& -\sum_{{\bf p_2p_3p_4}} \langle \bp_1 , \bp_2
\vert W\vert \bp_3 , \bp_4 \rangle 
(2\pi) \delta(\ep_{p_1}+\ep_{p_2}-\ep_{p_3}-\ep_{p_4}) \nonumber \\
&\times&
\Bigl\{ f(\bp_2)\, [1-f(\bp_3)]\,  [1-f(\bp_4)] 
+  [1-f(\bp_2)]\, f(\bp_3)  \, f(\bp_4) \Bigr\},
\end{eqnarray}
which gives
\be\label{I1bis}
I^{(1)}(\bp_1) &=&
\sum_{ {\bf p}_2  {\bf p}_3 {\bf p}_4 }\sum_{   {\bf p}_5 \, {\bf  p}_6\, {\bf p}_7}
\langle \bp_1 , \bp_2\vert W\vert \bp_3 , \bp_4 \rangle 
\frac{{\cal    P'}}{\ep_{p_1}+\ep_{p_2}-\ep_{p_3}-\ep_{p_4}} \nonumber \\
&\times &
\Big\vert\Big\langle  {(\bp_1- \bp_5)/}{2} \Big\vert T^{R}(\bp_1+\bp_5, 
\ep_{p_3}+\ep_{p_4}+\ep_{p_5}-\ep_{p_2}, x) \Big\vert
{(\bp_6-\bp_7)/}{2}\Big\rangle\Big\vert ^2  \nonumber \\&\times&
(2\pi)^4\, \delta(\bp_3+\bp_4+\bp_5-\bp_2-\bp_6-\bp_7)
~\delta(\ep_{p_3}+\ep_{p_4}+\ep_{p_5}-\ep_{p_2}-\ep_{p_6}-\ep_{p_7})\nonumber \\
&\times &
\Bigl\{f(\ep_{p_3}+\ep_{p_4}-\ep_{p_2})\, f(\bp_2)
\left[1-f(\bp_3)\right]\left[1-f(\bp_4)\right]\nonumber \\
&-&\left[1-f(\ep_{p_3}+\ep_{p_4}-\ep_{p_2})\right]
\left[1-f(\bp_2)\right]f(\bp_3)\, f(\bp_4)\Bigr\} 
\nonumber \\
&\times&\Bigl\{f(\bp_5)\left[1-f(\bp_6)\right]
\left[1-f(\bp_7)\right] +\left[1-f(\bp_5)\right]
f(\bp_6)\, f(\bp_7)\Bigr\} .  
\ee 
This result is essentially self-explanatory; the underlying process
is  illustrated in Fig. 2 b. 

Let us turn to the collision term in the kinetic equation (\ref{KINEQ4}).
The first term in the virial expansion is beyond the quasiparticle 
approximation, and e.g. for  the collision term $I_t$ has
the form 
\be \label{It}
I_t^{(0)}(\bp) &=&\sum_{{\bf p}_2{\bf p}_3{\bf p}_4} \frac{{\cal P'}}{\ep+\ep_{p_2}-\ep_{p_3} -\ep_{p_4}}
\langle \bp , \bp_2 \vert W\vert \bp_3 , \bp_4 \rangle \nonumber \\
&\times&\Bigl\{f(\bp)\, f(\bp_2) [1-f(\bp_3)][1- f(\bp_4)]  
+ [1-f(\bp)][1- f(\bp_2)]  f(\bp_3)\, f(\bp_4)  \Bigr\}. 
\ee
Compared with the first term of the virial expansion of 
the regular collision integral $I(\bp,x)$ this term 
differs by the off-shell energy denominator; in contrast to the 
virial correction terms here all four particles are in 
the final on-shell states, however 
their energies are not matched by the energy conservation condition.
The latter expression defines the correlated part of the distribution 
function as a functional of the on-shell QP distribution function via 
integral form of the kinetic equation (\ref{KINEQ4}).

\section{Virial Corrections in Equilibrium}

When integrated over momentum the kinetic equation provides number 
density conservation; in the equilibrium limit the conserved density 
is the sum of  quasiparticle density and the density of correlated excitations
\be 
n_{\rm tot} (\br )=  n_{\rm free}(\br)
+n_{\rm corr}(\br )=\int\!\frac{d^3p}{(2\pi)^3}\left[f(\ep_p, \br )
+\{f(\ep_p,\br )\}_{\rm corr}\right]. 
\ee
Consistent with our approximations, the equation 
of state contains virial
corrections up to the third order. Before giving the expression for 
the third quantum virial coefficient let us briefly recapitulate 
the second virial coefficient (see also ~\cite{KMGR}). 

\subsection{The Second Virial Coefficient}
The starting expression is that for the width at the 
two-body $T$-matrix level
\be 
\gamma(p_1) &=& -2i\sum_{p_2}\{ \langle (\bp_1-\bp_2)/2 \vert   
{\rm Im}T^{R}(p_1+p_2)\vert (\bp_1-\bp_2)/2 \rangle \, G^{<}(p_2)\nonumber \\
&-&\langle {(\bp_1 -\bp_2)/}{2}\vert T^{<}(p_1+p_2)
\vert (\bp_1-\bp_2)/{2}\rangle\, {\rm Im} G^{R}(p_2)
\}.
\ee
First we simplify this expression by eliminating the 
$T^<$ function using the optical theorem
(\ref{TMIN}) in equilibrium. The latter can be transformed,
using the operator relation $2\, {\rm Im} T^R =i( T^{<}-T^{>})$,
to the following form:
\be\label{OPT4} 
\langle \bp_1 \vert {\rm Im}T^{R}(P)
\vert \bp_2 \rangle
&=& -\frac{1}{2}\sum_{p_3p_4}a(\omega_3)\, a(\omega_4)  
\langle \bp_1 \vert T^{R}(P)
\vert {(\bp_3-\bp_4)/}{2}\rangle\, 
\left[1-f(\omega_3)-f(\omega_4)  \right]\nonumber \\
&\times&\langle {(\bp_3-\bp_4)/}{2} \vert T^{A}(P)
\vert \bp_2 \rangle\, (2\pi)^4\delta(P-p_3-p_4),
\ee
where $f(\omega)=\left[{\rm exp}\left(\beta(\omega -\mu)\right)
+1\right]^{-1}$ 
is the Fermi distribution function, $\mu$ being the  
chemical potential,  and $\beta$ the inverse temperature. Introducing the 
distribution functions in the eq. (\ref{TMIN}), noting that in
equilibrium 
$$f(\omega_1)f(\omega_2) = g(\omega_1+\omega_2)\,
[1-f(\omega_1)-f(\omega_2)], $$ where 
$g(\omega)=\left[{\rm exp}\left(\beta(\omega -\mu)\right)
-1\right]^{-1}$      
is the Bose distribution function, and comparing eqs. (\ref{TMIN}) and 
(\ref{OPT4}) one finds that 
\be 
\langle \bp_1 \vert T^{<}(P)\vert \bp_2 \rangle = 
-2i\, g(\omega_1+\omega_2)\, 
\langle \bp_1 \vert {\rm Im}T^{R}(P)\vert \bp_2 \rangle .
\ee
The expression for the width $\gamma(p)$ simplifies to 
\be 
\gamma(p_1) = 2 \sum_{p_2} a(\omega_2)\, 
\left[g(\omega_1+\omega_2)+f(\omega_2)\right]
\langle {(\bp_1-\bp_2)/}{2} \vert {\rm Im}T^{R}(p_1+p_2)
\vert {(\bp_1-\bp_2)/}{2}\rangle . 
\ee
Substituting the expression for $\gamma(p)$, eq. (\ref{GQPA}), 
in the expression for the correlated density 
and using  the equilibrium identity
\be 
\left[g(E)+f(\omega_2)\right]\left[f(E-\omega_2)-f(\omega_1)\right] 
=\left[g(E)-g(\omega_1+\omega_2)\right]
\left[1-f(\omega_1)-f(\omega_2)\right], 
\ee
one finds the following form of the two-particle correlation 
function~\cite{ZIM,KKL,SRS}
\be 
\{f^{II}(\ep_{p_1})\}_{\rm corr} 
= \int\!\frac{dE}{2\pi} g(E) b(\ep_{p_1}, E),
\ee
where 
\be 
b(\ep_{p_1}, E)& =& 2\sum_{{\bf p}_2}
\left[1-f(\ep_{p_1})-f(\ep_{p_2})\right]   
\{\langle (\bp_1-\bp_2)/2 \vert    
{\rm Im}T^{R}(E)\vert (\bp_1-\bp_2)/2 \rangle 
\frac{d}{d E}{\rm Re}{R}^{R}(E)\nonumber \\  
&+&{\rm Im}{R}^{R}(E)\, \frac{d}{d E} 
\langle (\bp_1-\bp_2)/2 \vert{\rm Im}T^{R}(E) 
\vert (\bp_1-\bp_2)/2  \rangle \}
\ee
is the two-particle spectral function and $R$ is the two-particle resolvent.

\subsection{The Third Virial Coefficient}

To derive the third virial coefficient we first relate the ${\cal T^{<,>}}$ 
matrices to the imaginary part of the retarded ${\cal T}^R$- matrix.
The optical theorem reads 
\be
\langle \bk_{23} \bq_1 \vert{\rm Im}{\cal T}^{R}
(K)\vert \bk_{45} \bq_6 \rangle 
&=& \frac{i}{2} \sum_{\alpha\beta}\sum_{p_4,p_5,p_6}
\langle \bk_{23} \bq_1 \vert {\cal T}^{R\,(\alpha)}(K)
\vert \bk_{45} \bq_6 \rangle 
\Bigl[G^<(p_4) G^<(p_5) G^<(p_6) \nonumber \\
&-& G^>(p_4) G^>(p_5) G^>(p_6) \Bigr]
\langle \bk_{45} \bq_6 \vert {\cal T}^{A\,(\beta)}(K)
\vert \bk_{23} \bq_1 \rangle \nonumber 
\\
&\times&\delta^4\left(K-p_4-p_5-p_6\right).
\ee
where the momenta $p_i$ $i = 1\dots 6$ are related to $k_{\alpha\beta}, q_{\gamma}$
via relations (\ref{JACOBI}).
Using the equilibrium 
identity
\be 
f(\omega_1) f(\omega_2)  f(\omega_3) & = &
f(\omega_1+\omega_2+\omega_3)\Bigl[1-f(\omega_1)-f(\omega_2)-f(\omega_3)
+f(\omega_1)f(\omega_2)\nonumber \\
&+&f(\omega_1)f(\omega_3)+f(\omega_2)f(\omega_3)
 \Bigr],
\ee
one finds in the QP limit:
\be 
\langle \bk_{23} \bq_1 \vert {\cal T}^{< }(K)\vert \bk_{45} \bq_6 \rangle
&=& 2i f\left(\omega_1+\omega_2 + \omega_3\right)
\langle \bk_{23} \bq_1 \vert{\rm Im}{\cal T}^{R}(K)
\vert \bk_{45} \bq_6 \rangle ,
\nonumber \\
\langle \bk_{23} \bq_1 \vert {\cal T}^{>}(K)\vert \bk_{45} \bq_6 \rangle
&=& 2i\left[ 1- f\left(\omega_1+\omega_2 + \omega_3\right) \right]
\langle \bk_{23} \bq_1 \vert{\rm Im}{\cal T}^{R}
(K)\vert \bk_{45} \bq_6 \rangle .
\ee
The three-particle correlation function takes the form 
\be\label{V3}
\{ f^{III}(\ep_{p_1})\}_{\rm corr} &=& - 2 i 
\int\!\frac{d\omega}{2\pi}
\sum_{p_2,p_3} \langle \bk_{23} \bq_1 \vert{\rm Im}{\cal T}^{R}(K)
\vert \bk_{23} \bq_1 \rangle \left[f \left( \omega_1 + \ep_{p_2} +\ep_{p_3}
\right) + g(\ep_{p_2} +\ep_{p_3})\right] \nonumber \\
&\times& \left[1- f(\ep_{p_2}) - f(\ep_{p_3}) \right]\, 
\left[f(\omega) - f(\ep_{p_1})\right]
\frac{{\cal P'}}{\ep_{p_1}-\omega}.
\ee
Next we use the identity, valid in equilibrium, 
\be 
&&\left[ f(\omega+\ep_{p_2}+\ep_{p_3}) + g(\ep_{p_2}+\ep_{p_3})\right]
\left[ 1- f(\ep_{p_2}) -f(\ep_{p_3})\right] \left[f(\omega)-f(\ep_{p_1})
\right]\nonumber \\
&=&
\left[f(E)-f(\ep_{p_1}+\ep_{p_2}+\ep_{p_3}) \right] \Bigl[
1-f(\ep_{p_1}) -f(\ep_{p_2})- f(\ep_{p_3}) + f(\ep_{p_1}) f(\ep_{p_2})\nonumber \\
&+&f(\ep_{p_1}) f(\ep_{p_3}) +f(\ep_{p_2}) f(\ep_{p_3})\Bigr],
\ee
where $E = \omega + \ep_{p_2} +\ep_{p_3} $. After the substitution 
in eq. (\ref{V3}) the desired form of the correlation function is 
found
\be 
\{f^{III}(\ep_{p_1})\}_{\rm corr} = \int\!\frac{dE}{2\pi} f(E) ~c(\ep_{p_1}, E),
\ee
where the third order quantum virial coefficient is defined as 
\be 
c(\ep_{p_1}, E)& =& 2\sum_{{\bf p}_2{\bf p}_3}
\Bigl[
1-f(\ep_{p_1}) -f(\ep_{p_2})- f(\ep_{p_3}) + f(\ep_{p_1}) f(\ep_{p_2})
+f(\ep_{p_1}) f(\ep_{p_3}) +f(\ep_{p_2}) f(\ep_{p_3})\Bigr]\nonumber  \\
&\times&
\left[ \langle \bk_{23} \bq_1 \vert {\rm Im}{\cal T}^{R}
(K)\vert \bk_{23} \bq_1 \rangle
\frac{d}{d E} {\rm Re} {\cal R}(E)+ {\rm Im} {\cal R}(E) \frac{d}{d E} 
\langle \bk_{23} \bq_1 \vert {\rm Re} {\cal T}
(K)\vert \bk_{23} \bq_1 \rangle
\right]
\ee
and where ${\cal R} = \left[E- \ep_{p_1}-\ep_{p_2}-\ep_{p_3}
+i\eta\right]^{-1}$is the three-particle resolvent.

Collecting the results, the virial expansion for the density reads
\be 
n_{\rm tot} = \int\!\frac{d^3p dE}{(2\pi)^4} \left[ a(E,\ep_{p})~f(E) +   
b(E,\ep_{p})~ g(E) + c(E,\ep_{p})~ f(E) \right].
\ee
This expression extends the quantum virial expansion for equilibrium Fermi-systems 
to terms of the third order in density. We note that the 
last term, as in the two-body case~\cite{ZIM,KKL,SRS}, 
can be separated further into bound and scattering contributions,  
which can then be expressed through the in-medium three-body 
scattering phase-shifts.

\section{Summary}

In this paper, we derived coupled quantum kinetic equations for 
Fermi-systems in the framework of real-time Green's function theory. 
The quasiclassical double-time Kadanoff-Baym equation was reduced 
to two single-time Boltzmann-type equations for quasiparticles 
and off-mass-shell excitations which are coupled via generalized
scattering integrals. The part of correlations related to the 
 off-mass-shell excitations are accounted 
for via an expansion of the spectral function 
up to the  next-to-leading order terms in spectral width~\cite{SLM,SL},
which is regarded as a small parameter. Although the treatment of 
the drift part 
is conceptually close to refs. ~\cite{SLM,SL}, we show that a different 
partition of the quasiclassical functions with respect to the orders of the 
spectral width can be exploited with the same efficiency as the separation 
in  pole and off-pole contributions. In particular, 
the latter partition has the
advantage that it fulfills the spectral sum rule at each order of the decomposition 
and, in addition, closely parallels the equilibrium 
treatments based on the same 
approximation to the spectral function~\cite{ZIM,KKL,SRS}, 
thus providing a consistency check for the variables 
treated in the local equilibrium limit (e.g. scattering amplitudes).

The collision integrals contain contributions 
from  three-body scattering. 
The three-particle scattering amplitudes are 
determined by mapping the Faddeev 
decomposition on the time-contour. The resulting in-medium
three-body equations contain the effects of the statistical 
suppression of the 
intermediate state three-particle 
propagation and quasiparticle renormalization;
in the vacuum limit they reduce  to the well-known 
Faddeev equations for the three-particle 
problem in free space~\cite{FADDEEV}. 
The amplitudes derived in this manner differ
from those considered by Bezzerides and DuBois ~\cite{DUBOIS} in that 
they resum the complete series in the particle-particle channel and are 
free of divergences associated with the non-Fredholm nature of the kernel 
of the three-body integral equation for the amplitudes. 
 Our equations appear to be consistent with the 
time-local limit of the 2-particle--hole equations derived in the self-consistent
random-phase approximation~\cite{DAN,SCHUCK,SCHUCK2,DUKELSKY}.

The decomposition of the scattering integrals 
in the on- and off-shell parts is 
accomplished in much the same manner as for the scattering amplitudes. 
The zeroth order term is the familiar Landau collision integral in the 
$T$-matrix approximation~\cite{BP,KB,DAN}. Four further terms 
represent first order  virial corrections 
to this collision integral, with one of the incoming or 
outgoing particles being 
in an off-mass-shell state.  Consistent with keeping only the first order 
terms in the width, the transition probabilities enter the zeroth order 
term of the virial expansion with the full off-shell 
contribution, while the first order terms acquire only 
on-shell (zeroth order)
contributions. A similar expansion holds for the kinetic equation for 
correlated elementary excitations. The scattering processes
that contain an off-shell particle in 
the final state can further be expanded, 
putting all final states  on the mass-shell: one then finds that these
processes effectively correspond to successive two-body scattering events 
connected by an intermediate off-shell propagation. 

In the equilibrium limit the equation of state has 
the form of a virial expansion 
and is truncated at the level of the three-body 
correlation. An explicit expression 
for the third quantum virial coefficient, which 
is expressed via the three-body
scattering amplitudes and the three particle resolvent, is obtained. 
The latter result extends previous discussions of the quantum 
Beth-Uhlenbeck formula for the density including the two-body 
correlation~\cite{ZIM,KKL,SRS} to the three-body case.

Numerical studies of three-body correlations at finite temperatures and 
densities in the context of excited nuclear matter using the 
full three-body equations either in the AGS or Faddeev formulation 
are underway. The in-medium nucleon-deuteron cross-section has been calculated
recently~\cite{BRS}, and the relaxation times for the nucleon-deuteron 
system have been estimated~\cite{BR}. The Mott-dissociation of 
tritium in  hot nuclear matter is now being studied.

\section*{Acknowledgments} The authors acknowledge helpful conversations with 
M. Beyer, P. Lipavsk\' y, K. Morawetz, V. G. Morozov, V. \v Spi\v cka and 
D. N. Voskresensky. AS wishes to acknowledge the support of the 
Max Planck Society at the early stage of this work and  the Max Kade 
Foundation (New York, NY) for a fellowship during the completion of the manuscript.


\begin{thebibliography}{99}
\bibitem{BP}G. Baym and C. J. Pethick, 
``Landau Fermi-Liquid Theory'',  Wiley, New York, 1991.
\bibitem{LP} E. M. Lifshitz and L. P. Pitaevskii, {``Physical Kinetics''}, Pergamon, 
Oxford, 1981.
\bibitem{ZMR}D. N. Zubarev, V. G. Morozov, and G. R\"opke, { 
``Statistical Mechanics of Nonequilibrium Processes''}, vol. I, Akademie Verlag, Berlin, 1996.
\bibitem{BERTSCH} G. F. Bertsch and Das Gupta, {\it Phys. Rep.} {\bf 160} (1988), 185. 
\bibitem{BALDO} M. Baldo, U. Lombardo, and P. Schuck, {\it Phys. Rev. C} {\bf 52}   (1995), 975.
\bibitem{DB} P. Danielewicz and G. F. Bertsch, {\it Nucl. Phys. A}  {\bf 533}  (1991), 712.
\bibitem{RS} G. R\"opke and H. Schulz, {\it Nucl. Phys. A } {\bf 477}  (1988), 472.
\bibitem{VOSKRESENSKY} J. Knoll and D. N. Voskresensky, {\it Ann. Phys. (NY)} 
{\bf 249}  (1996), 532,  
and references therein.
\bibitem{COHEN} E. G. D. Cohen, {\it Physica A}  {\bf 194} (1993), 229.
\bibitem{SNYDER} R. F. Snyder, {\it Journ. Stat. Phys.} {\bf 61} (1990), 443.
{\it ibid.} {\bf 63}  (1991), 707.
\bibitem{LALOE} F. Lal\"oe and W. J. Mullin, {\it Journ. Stat. Phys.} {\bf 59}  (1990), 725.
\bibitem{MCLENNAN}  J. A. McLennan, {\it Journ. Stat. Phys.} {\bf 28}  (1982), 521.
\bibitem{KLIMONTOVICH} Y. L. Klimontovich, D. Kremp, and W. D. Kraeft, 
``Advances in Chemical Physics'', Wiley, New York, 1987.
\bibitem{KB}L. P. Kadanoff and G. Baym, ``Quantum Statistical Mechanics''
   Benjamin, New York,  1962.
\bibitem{KELDYSH} L. V. Keldysh, {\it Sov. Phys. JETP} {\bf 20}  (1965), 1018.
\bibitem{SERENE} J. W. Serene and D. Reiner, {\it Phys. Rep.} 
{\bf 101}  (1983), 221
\bibitem{DAN} P. Danielewicz, {\it Ann. Phys. (NY)}
{\bf 152}  (1984), 239; {\it ibid.} {\bf 197}  (1990), 154.
\bibitem{ARONOV} A. G. Aronov, Yu. M. Gal'perin, V. L. Gurevich and V. I. Kozub
{\it in} {``Nonequilibrium Superconductivity''} 
(D. N. Langenberg and A. I. Larkin, Eds.) p. 325, North Holland, Amsterdam, 1986. 
\bibitem{BMALF}{W. Botermans and R. Malfliet, {\it Phys. Rep.}
 {198} (1990), 115.}
\bibitem{KMGR} K. Morawetz and G. R\"opke, {\it Phys. Rev. E} {\bf 51}   (1995), 4246.
\bibitem{SLM}V. \v Spi\v cka and P. Lipavsk\' y, K. Morawetz, 
{\it Phys. Rev. B} {\bf 55}   (1997), 5095.
\bibitem{SL} V. \v Spi\v cka and P. Lipavsk\' y, {\it Phys. Rev B} {\bf 52}  (1995), 14615.
\bibitem{ZIM} R. Zimmermann and H. Stolz, {\it Phys. Status Solidi B} {\bf 131}
    (1985), 151;  {\it ibid.} {\bf 94},   (1979), 139.
\bibitem{KKL} D. Kremp, W. D. Kraeft, and A. J. M. D. Lambert, {\it Physica A} {\bf 127},
  (1984), 72.
\bibitem{SRS} M. Schmidt, G. R\" opke and H. Schulz, {\it Ann. Phys. (NY)}
    {\bf 202}  (1990), 57.
\bibitem{FADDEEV} L. D. Faddeev, {\it Sov. Phys. JETP} {\bf 39}  (1960), 139.
\bibitem{AGS} E. O. Alt, P. Grassberger, and W. Sandhas, {\it Nucl. Phys. B} {\bf  2}, 
 (1967), 167.
\bibitem{DUBOIS} B. Bezzerides and D. F. DuBois, {\it Phys. Rev.}  {\bf 186}  (1968), 233.
\bibitem{SCHUCK} P. Schuck, F. Villars, and P. Ring, {\it Nucl. Phys. A} {\bf 208}  (1973), 302.
\bibitem{SCHUCK2} P. Schuck, {\it Nucl. Phys. A } {\bf 567} (1994), 78.
\bibitem{DUKELSKY}
 J. Dukelsky and P. Schuck, {\it Nucl. Phys. A} {\bf  512}  (1990), 466.
\bibitem{DS} P. Danielewicz and P. Schuck, {\it Nucl. Phys. A} {\bf 566}  (1994), 23.
\bibitem{BRS} M. Beyer, G. R\"opke and A. Sedrakian, {\it Phys. Lett. B}  {\bf 376}  (1996), 7.
\bibitem{BR} M. Beyer and G. R\"opke, {\it Phys. Rev. C} {\bf 56} (1997) 2636.
\bibitem{BRUECKNER} K. A.  Brueckner, and J. L. Gammel, {\it Phys. Rev.} {\bf 109},  (1958), 1023.
\bibitem{GALITSKI} V. M. Galitskii, {\it Sov. Phys. JETP} {\bf 7}  (1958), 104.
\bibitem{THOULESS} D. J. Thouless, {\it Ann. Phys. (NY)} {\bf  10}  (1960), 553.
\bibitem{SAL} A. Sedrakian, Th. Alm and U. Lombardo, {\it Phys. Rev. C} {\bf 55} (1996), R582.
\bibitem{HEISELBERG} H. Heiselberg, C. J. Pethick 
and D. G. Ravenhall, {\it Ann. Phys. (NY)} {\bf 223}  (1993), 37.
\bibitem{GRKMTA} G. R\"opke, K. Morawetz and Th. Alm, 
{\it Phys. Lett. A}  {\bf 194} (1994), 1.
\bibitem{ROEPKE} G. R\"opke et al., {\it Nucl. Phys. A}  {\bf 399},  (1983) 587.
\bibitem{KREMP}  D. Kremp, M. Schlanges, Th Bornath, {\it Journ. Stat. Phys.} {\bf 41}
 (1985), 661.
\end{thebibliography}
\end{document}